\documentclass[12pt]{iopart}
\pdfoutput=1
\usepackage{iopams} 
\usepackage{graphicx}
\expandafter\let\csname equation*\endcsname\relax
\expandafter\let\csname endequation*\endcsname\relax
\usepackage{amsmath}

\begin{document}
\title[A state-insensitive nanofiber trap]{A state-insensitive, compensated nanofiber trap}
\author{C Lacro\^ute$^1$ \footnote[1]{These authors contributed equally to this work.}, K S Choi$^{1,2}$\footnotemark[1], A Goban$^{1}$\footnotemark[1], D J Alton$^1$, D Ding$^1$, N P Stern$^{1,3}$ and H J Kimble$^1$}
\address{$^1$ Norman Bridge Laboratory of Physics 12-33\\ California Institute of
Technology, Pasadena, California 91125, USA}
\address{$^2$ Spin Device Research Center \\ Korea Institute of Science and Technology \\
39-1 Hawolgok-dong, Seongbuk-gu, Seoul, 136-791, Korea}
\address{$^3$ Current address: Department of Physics and Astronomy \\ Northwestern University \\ 2145 Sheridan Rd., Evanston, IL 60208}

\begin{abstract}
Laser trapping and interfacing of laser-cooled atoms in an optical fiber network is an important capability for quantum information science. Following the pioneering work of Balykin \textit{et al.} and Vetsch \textit{et al.}, we propose a robust method of trapping single Cesium atoms with a two-color state-insensitive evanescent wave around a dielectric nanofiber. Specifically, we show that vector light shifts (i.e., effective inhomogeneous Zeeman broadening of the ground states) induced by the inherent ellipticity of the forward-propagating evanescent wave can be effectively canceled by a backward-propagating evanescent wave. Furthermore, by operating the trapping lasers at the magic wavelengths, we remove the differential scalar light shift between ground and excited states, thereby allowing for resonant driving of the optical $D_2$ transition. This scheme provides a promising approach to trap and probe neutral atoms with long trap and coherence lifetimes with realistic experimental parameters.
\end{abstract}
\maketitle

\section{Introduction}

The development of a matter-light quantum interface using cold atoms and optical fibers has been an active field of research over the past years \cite{Kimble2008}. Recent advances towards this goal include the observation of electromagnetically induced transparency and the loading of ultracold atoms in hollow-core optical fibers \cite{Renn1995,Ito1996,Christensen2008,Bajcsy2009}, as well as the trapping and probing of atomic ensembles via the evanescent fields surrounding tapered nanofibers \cite{Vetsch2010,Kien2004a, Nayak2007,Sague2007}. While prominent examples of off-resonant interaction between evanescent waves and matter have used a plane dielectric geometry for atom optics and interferometry \cite{Balykin1988, Cronin2009} as well as for surface traps of quantum degenerate gases \cite{Ovchinnikov1997, Rychtarik2004, Bakr2010}, recent progress of atom-light interactions with optical waveguides \cite{Christensen2008,Bajcsy2009, Kien2004a, Nayak2007,Sague2007} sets the stage for the fiber integration of free-space quantum systems in a quantum network via quantum-state transfer between matter and light \cite{ Boozer2007, Choi2008, Hammerer2010, Sangouard2011} and for strong coupling of single atoms and photons trapped near microcavities \cite{Aoki2006,Alton2011,Stern2011,Barclay2006,Colombe2007,Trupke2007}.  Furthermore, these effective 1-dimensional systems may be applied for investigating quantum many-body phenomena in low dimensions with long-range interactions mediated by the waveguide \cite{Zoubi2010, Kitagawa2011, Shen2005, Chang2007}.

One major drawback of many optical traps is that spatially inhomogeneous energy shifts $U(\mathbf{r})$ generally depend on the atomic electronic state, limiting long-lived trap and coherence times necessary for repeated coherent operations \cite{Corwin1999}. This is traditionally alleviated by constructing a \emph{state-insensitive} optical trap designed to decouple atomic transition frequencies from the spatially varying potential of each electronic state \cite{Ye2008}.  Specifically, at the ``magic'' wavelength conditions, the differential response of the dynamic scalar polarizabilities $\alpha^{(0)}(\omega)$ for the ground and excited states $\alpha^{(0)}_{|g\rangle},\alpha^{(0)}_{|e\rangle}$  at the optical frequency $\omega$ can be tailored such that both levels are perturbed identically with $\alpha^{(0)}_{|g\rangle}=\alpha^{(0)}_{|e\rangle}$. This leads to a vanishing differential atomic level shift $\delta U_{\mathrm{scalar}}=0$ \cite{Ye2008, Kimble1999,Katori1999,Ido2000,McKeever2003}. Differential shifts for the hyperfine ground states can be minimized by using far off-resonant beams, whereas Zeeman coherence can be conveniently protected by using linearly polarized light in which the vector light shifts are zero.

Although such magic wavelengths can be used for nanofiber traps \cite{Kien2005a}, the strongly guiding nature of the waveguide inevitably leads to non-negligible longitudinal electric fields $\mathbf{E}_z$ in the evanescent region, which are out of phase with the transverse field $\mathbf{E}_\perp = (\mathbf{E}_x, \mathbf{E}_y)$. Here, $z$ refers to the direction parallel to the fiber axis, while $x$ and $y$ are the coordinates perpendicular to the fiber axis. The resulting local polarization at location $\mathbf{r}$ is in general elliptical even for linearly polarized input beams, and induces vector shifts $U_{\mathrm{vector}}$. The differential vector shift $\delta U_{\mathrm{vector}}$ in turn manifests itself as a ``fictitious magnetic field'', leading to inhomogeneous Zeeman broadening \cite{Dupont-Roc1967}.  Furthermore, the spatially varying elliptical polarization of the evanescent field on a scale $\delta r<\lambda$ renders it difficult to cancel $\delta U_{\mathrm{vector}}$ using bias fields, resulting in increased heating rate \cite{Corwin1999} and limited coherence time \cite{Felinto2005}.    

Building upon the recent realization of a nanofiber trap as proposed in Ref. \cite{Balykin2004} and demonstrated in Refs. \cite{Vetsch2010, Dawkins2011}, we propose a promising strategy for a state-insensitive evanescent field trap. Differential scalar shifts $\delta U_{\mathrm{scalar}}$ between $|g\rangle$ and $|e\rangle$  are canceled using ``magic'' wavelength conditions. The inhomogeneous Zeeman broadening $\delta U_{\mathrm{vector}}$ caused by a forward propagating blue-detuned field $\mathbf{E}^{\textrm{(fwd)}}$ is canceled by a backward propagating field $\mathbf{E}^{\mathrm{(bwd)}}$ with a small relative frequency detuning $\delta_{fb}$. Thus, our scheme can compensate for the light shifts of the strongly guided evanescent waves to the first order in the space external to the dielectric fiber, leading to favorable parameters for the realization of a long-lived fiber-integrated quantum memory and resonant coupling to ultra-high quality micro-cavities based on optically trapped atoms.

\section{A state-insensitive nanofiber trap}
In this section, we discuss an \emph{ab initio} calculation of the optical nanofiber trap for atomic Cesium. We show that the light shifts caused by the elliptically polarized components of the fiber's evanescent field are not negligible. We then propose a scheme to cancel these shifts and generate a two-color, state-insensitive, three-dimensional trap for Cs atoms along the nanofiber.

\subsection{ac Stark shift Hamiltonian}

We start by considering the Hamiltonian for an atom interacting with an electric field $\mathbf{E}$ in the dipole approximation: 
\begin{equation}
\begin{array}{llll}
\hat{H}_{\mathrm{ls}}&=-\mathbf{\hat{d}}\cdot\mathbf{\hat{E}},
\end{array}
\label{eq:dipole}
\end{equation}
where $\mathbf{\hat{d}}$ is the electric dipole operator and $\mathbf{\hat{E}}$ is the electric field operator. Taking into account the atomic hyperfine structure, this Hamiltonian can be decomposed into its spherical tensor components parameterized by the dynamic polarizability $\alpha(\omega)$ \cite{Deutsch2010a, Geremia2006,Deutsch1998}:
\begin{equation}
\begin{array}{llll}
\hat{H}_{\mathrm{ls}}&= \hat{H}_0+\hat{H}_1+\hat{H}_2
\\
&=\alpha^{(0)}\mathbf{\hat{E}^{(-)}}\cdot \mathbf{\hat{E}^{(+)}}
\\
&+i\alpha^{(1)}\frac{\left(\mathbf{\hat{E}^{(-)}}\times \mathbf{\hat{E}^{(+)}} \right) \cdot \mathbf{\hat{F}}}{F}
\\
&+\sum\limits_{\mu,\nu}\alpha^{(2)}\hat{E}^{(-)}_{\mu}\hat{E}^{(+)}_{\nu}\frac{3}{F(2F-1)}\left [\frac{1}{2}(\hat{F}_{\mu}\hat{F}_{\nu}+\hat{F}_{\nu}\hat{F}_{\mu})-\frac{1}{3}\hat{F}^2\delta_{\mu\nu}\right],
\end{array}
\label{eq:hamiltonian}
\end{equation}
where $\alpha^{(0)}$, $\alpha^{(1)}$ and $\alpha^{(2)}$ are the scalar, vector and tensor atomic dynamic polarizabilities, $\mathbf{\hat{E}^{(+)}}$ and $\mathbf{\hat{E}^{(-)}}$ are the positive and negative frequency components of the electric field, $\mathbf{\hat{F}}=\mathbf{\hat{I}}+\mathbf{\hat{J}}$ is the atomic total angular momentum operator, with $\mathbf{\hat{I}}$ and $\mathbf{\hat{J}}$ the nuclear and electronic angular momentum operators, $\mu,\nu\in\{-1,0,1\}$ are components in the spherical tensor basis, and $\hat{H}_0$, $\hat{H}_1$, and $\hat{H}_2$ are the terms associated with the scalar, vector, and tensor light shifts, respectively. The light shifts $U_{\mathrm{scalar}}$, $U_{\mathrm{vector }}$, and $U_{\mathrm{tensor}}$ arising from each term have been expressed explicitly in Refs. \cite{Deutsch2010a, Geremia2006}.

For two-level atoms with ground and excited states $|g\rangle,|e\rangle$, the scalar shift $U_{\mathrm{scalar}}$ can be approximated by $U_{\mathrm{scalar}}\propto |\mathbf{E}|^2/\delta$ for detunings $\delta=\omega-\omega_a$ large compared to the excited state decay rate $\Gamma$, where $\omega$ is the electric field angular frequency and $\omega_a$ is the $|g\rangle\rightarrow|e\rangle$ transition frequency. The ground state will experience a repulsive potential for blue-detuned ($\delta > 0$) electric fields, and an attractive potential for red-detuned ($\delta < 0$) electric fields. The scalar dynamic polarizability $\alpha^{(0)}$ is in general different for the states $|g\rangle$ and $|e\rangle$ resulting in a differential scalar shift and a mismatch of the ground and excited state potentials. For the typically anti-trapped excited state, near-resonant driving of the transition by an additional beam with frequency $\omega_2 \simeq\omega_a$ can cause significant heating of a trapped atom \cite{Corwin1999}. This situation can be remedied by the use of ``magic'' wavelengths for which $\alpha^{(0)}_{|g\rangle}=\alpha^{(0)}_{|e\rangle}$ \cite{Ye2008,Kimble1999,Katori1999,McKeever2003}.

The vector term $\hat{H}_1$ of Eq. \eqref{eq:hamiltonian} induces a Zeeman-like splitting proportional to a projection of the total atomic angular momentum $\mathbf{F}$ and arises from a so-called ``fictitious magnetic field'' proportional to the ellipticity of the electric field \cite{Dupont-Roc1967}. In the case of  a free-space plane wave propagating along the $z$ axis, $\hat{H}_1$ can be expressed in terms of the Stokes operators $\mathbf{\hat{S}}=(\hat{S_0},\hat{S_x},\hat{S_y},\hat{S_z})$ as \cite{Geremia2006}:
\begin{equation}
\hat{H}_{1}\propto\alpha^{(1)}(\omega)\epsilon\frac{\hat{F_z}}{F},
\label{eq:VectorShift}
\end{equation}
where $\epsilon=\langle\hat{S}_z\rangle/\langle\hat{S}_0\rangle =\frac{|E_{+1}|^2-|E_{-1}|^2}{|E_{+1}|^2+|E_{-1}|^2}$ is the ellipticity of the electric field. For an elliptically polarized beam, the vector shift can be as large as the scalar shift, and can, for example, be used to cancel the differential light shifts of Rubidium atoms confined in a 3D optical lattice \cite{Chicireanu2011}.

The last term $\hat{H}_2$ in Eq. \eqref{eq:hamiltonian} represents the tensor shift. It vanishes for atoms with total angular momentum $F=1/2$ \cite{Geremia2006}. In the case of the $D_2$ transition of Cs, that we consider here, it will depend only on the electronic angular momentum $\mathbf{\hat{J}}$ for detunings large compared to the $6P_{3/2}$ excited state hyperfine structure, and vanish for $J=\frac{1}{2}$ \cite{Deutsch2010a, Geremia2006}. It will therefore only act on the excited state of the Cs $D_2$ transition, inducing shifts on the Zeeman $m_{F'}$ sublevels proportional to $m_{F'}^2$.

\subsection{Evanescent optical traps using the fundamental mode of the waveguide}
\begin{figure}[h!]
\centering
\begin{minipage}[c]{0.45\textwidth}
\includegraphics[width=\textwidth]{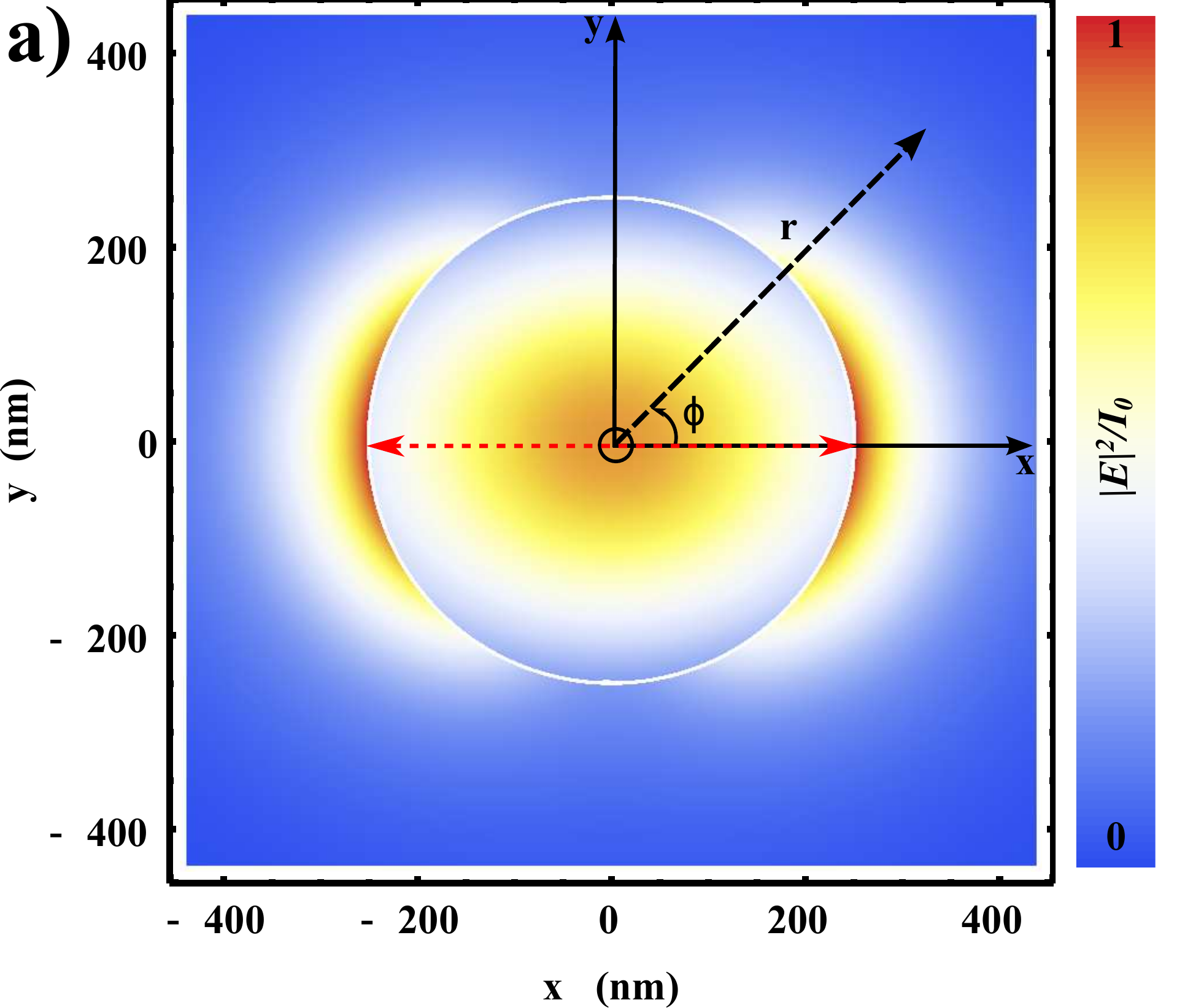}
\end{minipage}
\centering
\begin{minipage}[c]{0.4\textwidth}
\includegraphics[width=\textwidth]{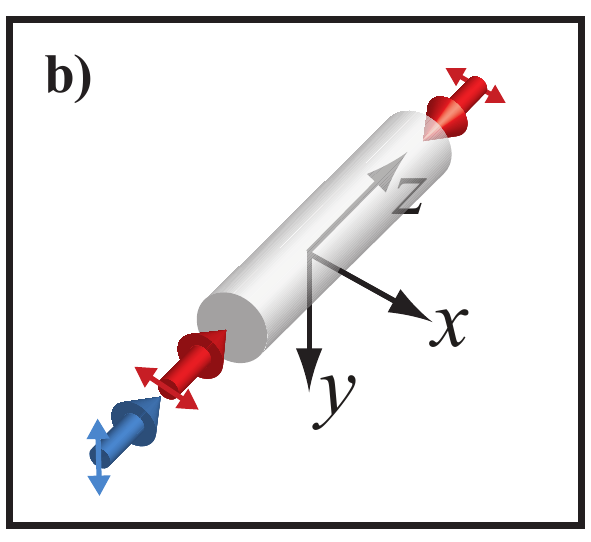}
\end{minipage}
\centering
\begin{minipage}[c]{0.4\textwidth}
\includegraphics[width=\textwidth]{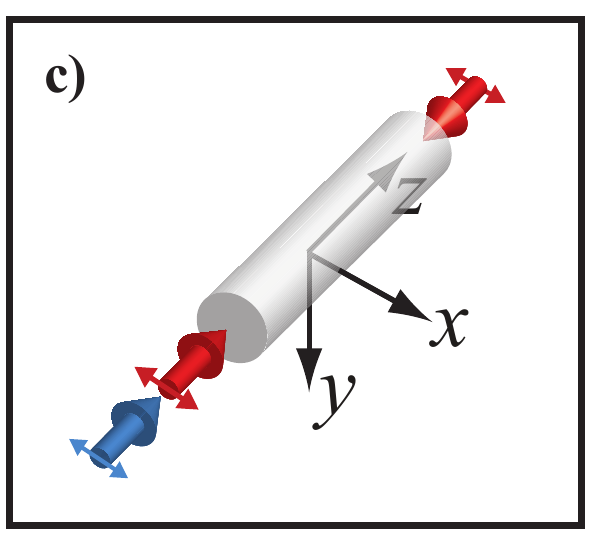}
\end{minipage}
\hspace{.04\textwidth}
\centering
\begin{minipage}[c]{0.4\textwidth}
\includegraphics[width=\textwidth]{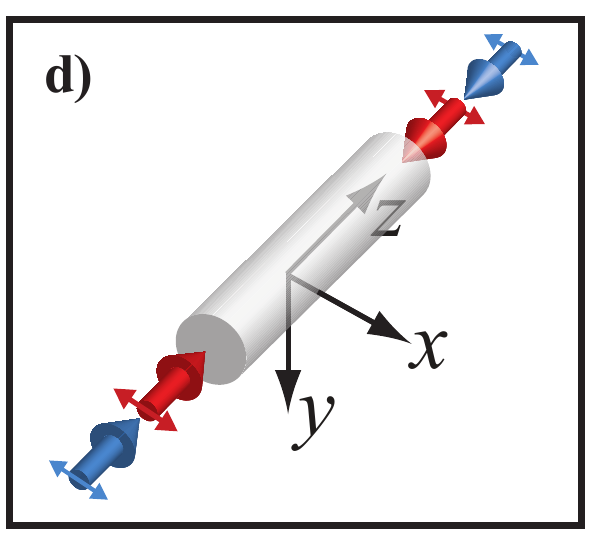}
\end{minipage}
\caption{Trapping schemes. \textbf{a)} Field intensity $|E|^2$ in the plane transverse to the fiber for a single, $x$-polarized beam at $\lambda=937\ \mathrm{nm}$. $|E|^2$ is normalized to the intensity just outside the fiber $I_0=|E(r=a_+, \phi=0)|^2$, with $a=250\ \mathrm{nm}$ and $a_+=a+|r-a|\ ,\ r \to a$. The red dashed arrow indicates the input polarization. \textbf{b)} Trapping scheme used in Ref. \cite{Vetsch2010}. Red(blue)-detuned beams are shown by red(blue) thick arrows. Input polarizations are shown by the thin arrows. A single, $y$-polarized blue-detuned beam is used. \textbf{c)} Three-beam scheme with parallel $x$-polarizations. All beams have an intensity maximum in the $x-y$ plane along the direction of the input polarization. \textbf{d)} A second $x$-polarized blue-detuned beam is added to compensate for the vector shifts, as discussed in the main text.}
\label{fig:configs}%
\end{figure}

When the radius $a$ of an optical fiber is reduced well below the propagating field wavelength $\lambda$, the resulting cladding-to-air waveguide supports only the ``hybrid'' fundamental mode $\mathrm{HE}_{11}$ \cite{Kien2004a, Tong2004}. In this strongly guiding regime, a significant fraction of energy of the $\mathrm{HE}_{11}$ mode is carried in the form of an evanescent wave outside of the nanofiber. The evanescent field intensity is azimuthally asymmetric when the input polarization is linear \cite{Kien2004a,Tong2004}. Fig. \ref{fig:configs}a shows the electric field intensity $|E|^2=|E_x|^2+|E_y|^2+|E_z|^2$ in a plane transverse to the fiber for a single, linearly-polarized beam. The unit vectors $(\mathbf{e}_x,\mathbf{e}_y,\mathbf{e}_z)$ form the basis of the ($x$,$y$,$z$) frame, and $(r,\phi)$ are the cylindrical coordinates in the transverse plane ($x$,$y$).

By appropriately combining blue-detuned and red-detuned fields $\mathbf{E}_{\mathrm{red}}$ and $\mathbf{E}_{\mathrm{blue}}$ in an optical nanofiber, an atomic trapping potential can be engineered from the evanescent electric fields \cite{Balykin2004}. Different configurations can be used for obtaining 3D confinement.
Here, we consider the schemes illustrated in Fig. \ref{fig:configs}, for an infinite SiO$_2$ cylindrical waveguide of radius $a=250 \ \mathrm{nm}$. In all configurations, the beams are linearly polarized at the waveguide input to ensure azimuthal confinement for trapped atoms. A pair of $x$-polarized red-detuned beams generates a 1D lattice along the fiber axis for longitudinal confinement. Fig. \ref{fig:configs}b shows the configuration used in Ref. \cite{Vetsch2010}, namely a pair of red-detuned, $x$-polarized beams, and a single blue-detuned, $y$-polarized beam. An alternative scheme would be to use the three beams with parallel polarizations, as illustrated in Fig. \ref{fig:configs}c. This scheme allows for the use of lower power for the blue-detuned beam, by about a factor of three for the parameters that we will consider in section \ref{sec:results}, but results in larger vector shifts as we will discuss in the next sections. The scheme that we propose makes use of four beams with parallel linear input polarizations, as shown in Fig. \ref{fig:configs}d. The additional blue-detuned beam compensates for the vector shifts of its companion blue-detuned beam, as we will show.

\begin{figure}[h!]\centering%
\includegraphics[width=0.95\columnwidth]{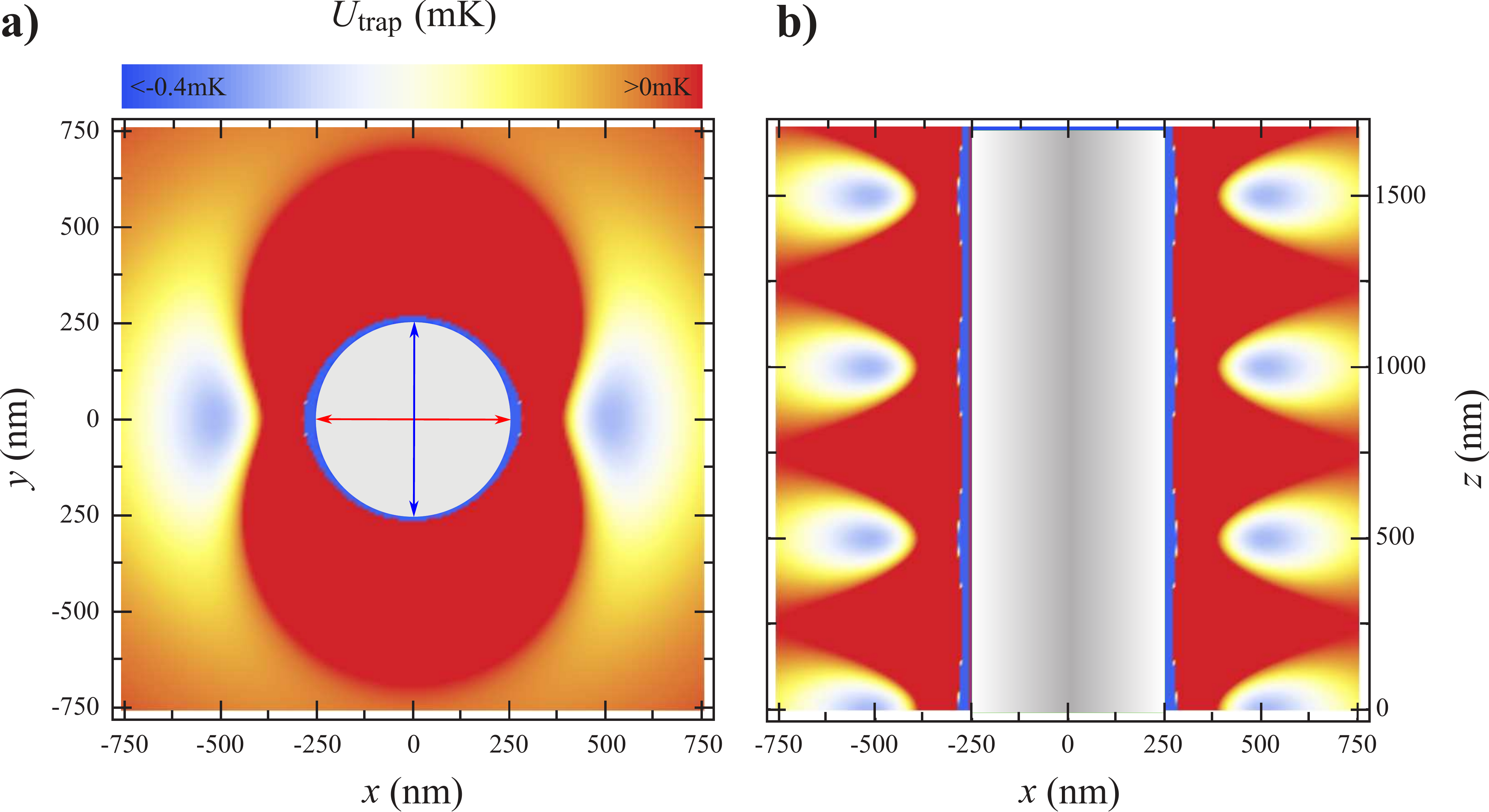}%
\caption{A ground state trapping potential $U_{\mathrm{trap}}$ is generated by two orthogonally polarized evanescent fields, confining Cs atoms outside a $500$ nm diameter optical fiber. Input polarizations are denoted by the arrows. \textbf{a)} $x$-$y$ plane. \textbf{b)} $x$-$z$ plane. $U_{\mathrm{trap}}$ results from two counter-propagating red-detuned beams ($1064$ nm, $2 \times 2.2$ mW), and a single blue-detuned beam ($780$ nm, $25$ mW), as shown in Fig. \ref{fig:configs}b \cite{Vetsch2010}. The standing wave structure of the attractive red-detuned field and the repulsive force from the blue-detuned beam enable 3D confinement of Cs atoms at each minimum of $U_{\mathrm{trap}}$ near the dielectric waveguide. $U_{\mathrm{trap}}$ diverges as the surface is approached due to the attractive van der Walls force.}
\label{fig:trap_vetsch}%
\end{figure}

Fig. \ref{fig:trap_vetsch} illustrates a trap generated using the configuration of Fig. \ref{fig:configs}b and the parameters in Ref. \cite{Vetsch2010}. About 2000 atoms were trapped in a 1-D lattice with a $50 \ \mathrm{ms}$ lifetime. This advance in the group of A. Rauschenbeutel represents an important milestone towards the micro-manipulation of ultra-cold atoms using evanescent field traps.

\subsection{$\mathrm{HE}_{11}$ mode - electric field polarization} \label{sub:comp}

The fundamental mode $\mathrm{HE}_{11}$ is often referred to as ``quasi-linear'' when excited with a linearly polarized input beam. However, for a dielectric waveguide in the strong-guiding regime with indices of refraction $n_1\approx1.5$ inside the waveguide and $n_2\approx1.0$ outside, the $\mathrm{HE}_{11}$ mode actually exhibits a significant ellipticity for $a\lesssim\lambda/2$, leading to vector shifts of the Zeeman sublevels. Formally, for a linearly polarized input, the evanescent field $\mathbf{E}=(E_x,E_y,E_z)$ of the fundamental mode propagating in the fiber can be expressed as follows for $r\geq a$ \cite{Kien2004a,Tong2004,Sague2008}:

\begin{subequations}
\label{eq:FundMode}
\begin{eqnarray}
\fl E_{x}(r,\phi,z,t) & = & A_{\mathrm{lin}}\frac{\beta_{11} J_1(h_{11} a)}{2q_{11}K_1(q_{11} a)}\label{eq:FundModea}\\
&&\times[(1-s_{11})K_0(q_{11} r) \cos(\varphi_0)+(1+s_{11})K_2(q_{11} r) \cos(2\phi-\varphi_0)]e^{i(\omega t-\beta_{11}z)}, \notag \\
\fl E_{y}(r,\phi,z,t) & = & A_{\mathrm{lin}}\frac{\beta_{11} J_1(h_{11} a)}{2q_{11}K_1(q_{11} a)}\label{eq:FundModeb}\\
&&\times[(1-s_{11})K_0(q_{11} r) \sin(\varphi_0)+(1+s_{11})K_2(q_{11} r) \sin(2\phi-\varphi_0)]e^{i(\omega t-\beta_{11}z)}, \notag \\
\fl E_{z}(r,\phi,z,t) & = & i A_{\mathrm{lin}}\frac{J_1(h_{11} a)}{K_1(q_{11} a)}K_1(q_{11} r) \cos(\phi-\varphi_0) e^{i(\omega t-\beta_{11}z)},\label{eq:FundModec}
\end{eqnarray}
\end{subequations}

with

\begin{subequations}
\label{eq:notations}
\begin{eqnarray}
s_{11} & = & \left[\frac{1}{\left(h_{11}a\right)^2}+\frac{1}{\left(q_{11}a\right)^2}\right] \left[\frac{J'_1(h_{11}a)}{h_{11}a J_1(h_{11}a)}+\frac{K'_1(q_{11}a)}{q_{11}a K_1(q_{11}a)}\right],\\
h_{11} & = & \sqrt{k_0^2 n_1^2 - \beta_{11}^2},\\
q_{11} & = & \sqrt{\beta_{11}^2 - k_0^2 n_2^2}.
\end{eqnarray}
\end{subequations}

Here, $\phi$ denotes the azimuthal position in the transverse plane (Fig. \ref{fig:configs}a), $\varphi_0$ indicates the polarization axis for the input polarization relative to the $x$ axis, $n_1$ and $n_2$ are the indices of refraction inside and outside the waveguide, $\beta_{11}$ is the mode propagation constant, $1/h_{11}$ is the characteristic decay length for the guided mode inside the fiber, $1/q_{11}$ is the characteristic decay length for the guided mode outside the fiber, $A_{\text{lin}}$ is the real-valued amplitude for the linearly polarized input, $J_{l}$ is the $l$-th Bessel function of the first kind, and $K_{l}$ is the $l$-th modified Bessel function of the second kind.

It is clear from Eq. \eqref{eq:FundMode} that the electric field intensity is not azimuthally symmetric. For a beam polarized along $\mathbf{e}_x$, i.e. $\varphi_0=0$, the intensity at the fiber's outer surface is maximum for $\phi=0,\ \pi$ and minimum for $\phi=\pm \pi/2$.

\begin{figure}%
\centering
\begin{minipage}[c]{0.3\textwidth}
\includegraphics[width=\textwidth]{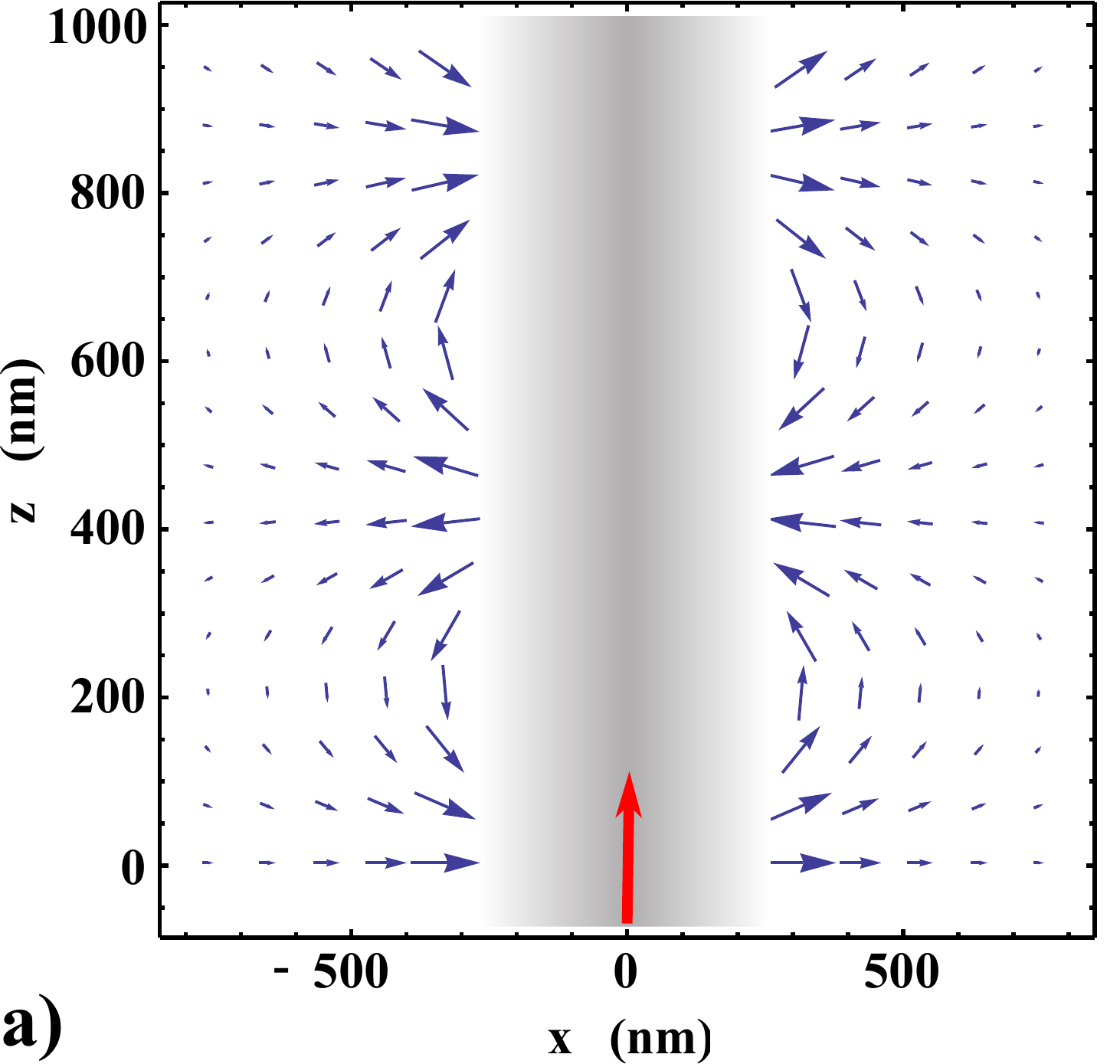}%
\end{minipage}
\hspace{0.02\textwidth}
\centering
\begin{minipage}[c]{0.3\textwidth}
\includegraphics[width=\textwidth]{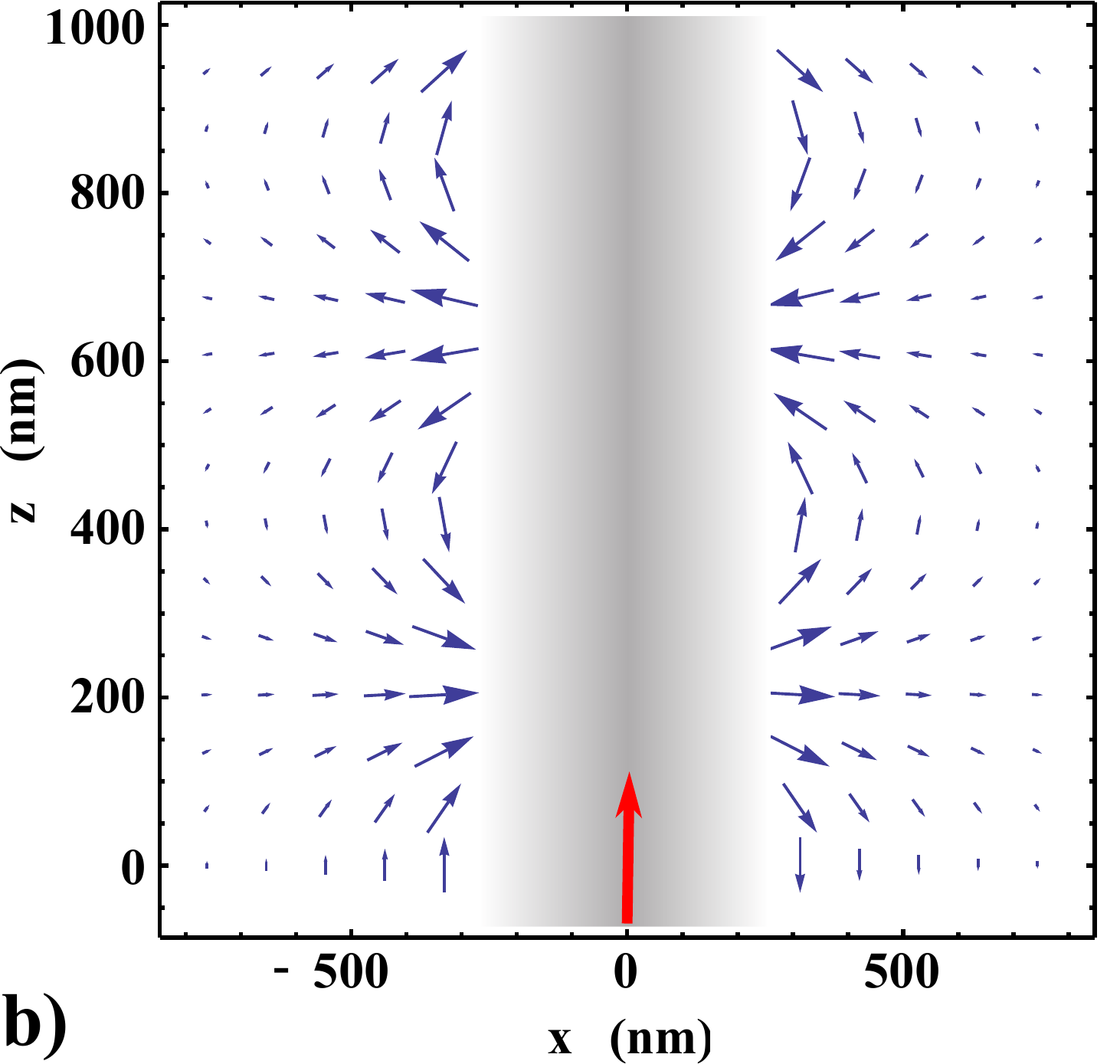}%
\end{minipage}
\hspace{0.02\textwidth}
\centering
\begin{minipage}[c]{0.3\textwidth}
\includegraphics[width=\textwidth]{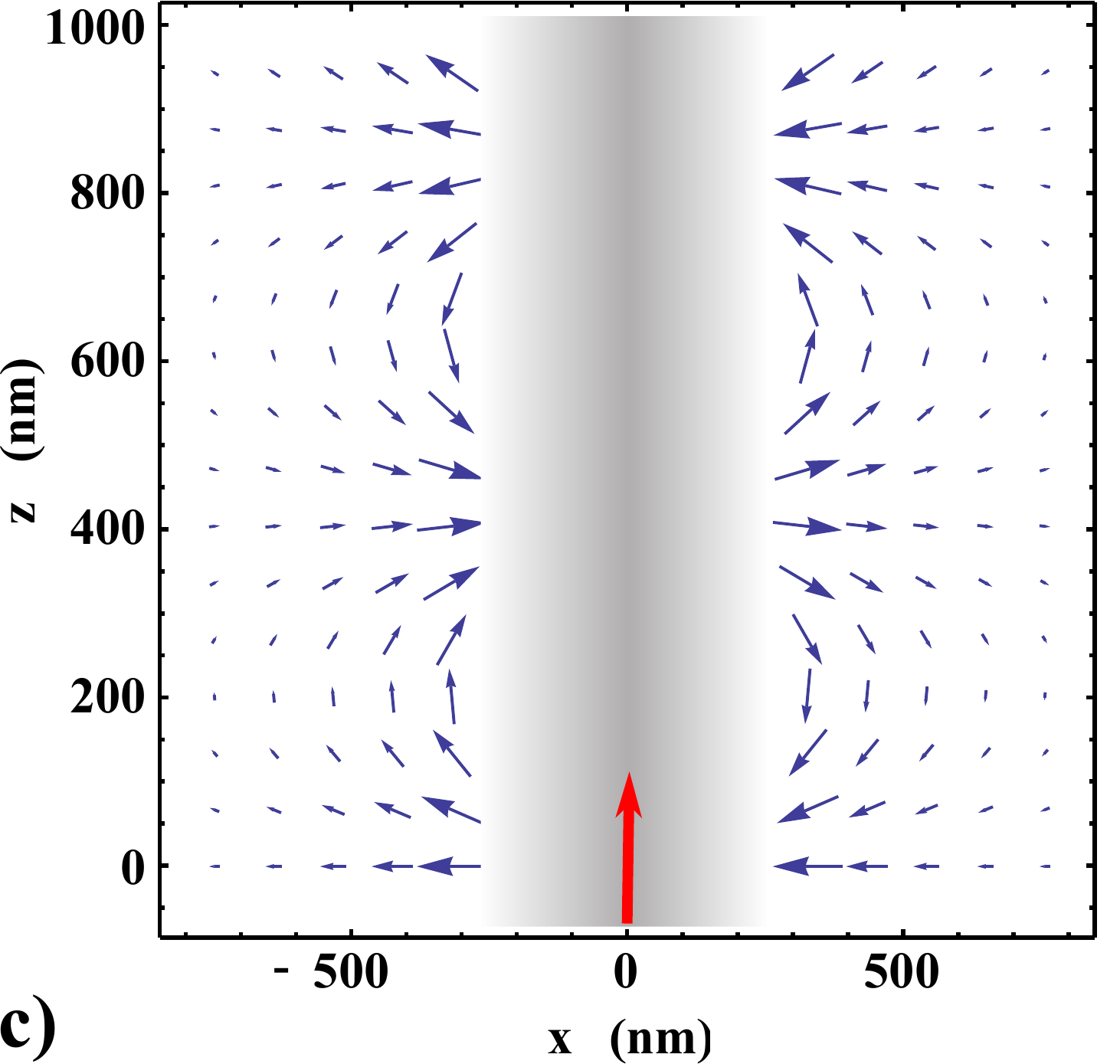}%
\end{minipage}
\caption{Electric field $\mathbf{E}(x,y,z,t)$ of a single propagating beam in the plane $y=0$. The input beam is $x$-polarized. The electric field $\mathrm{Re}[\mathbf{E}(x,y,z,t)]$, with $\mathbf{E}(x,y,z,t)$ defined as in Eq. \ref{eq:FundMode}, is shown by the blue arrows. The red arrow indicates the beam propagation direction. The field is shown for \textbf{a)} $\omega t=0$, \textbf{b)} $\omega t=\pi/2$, and \textbf{c)} $\omega t=\pi$.}%
\label{fig:sgle_polar}%
\end{figure}

Notably, the evanescent modes of the nanofiber have a significant longitudinal component $E_z$ along the fiber propagation direction, which is $\pi/2$ out-of-phase with the transverse components $(E_x,E_y)$ (Eq. \eqref{eq:FundModec}). At the outer fiber surface, $E_z$ is maximum for $\phi=\varphi_0,\varphi_0+\pi$ (i.e., along the input polarization axis), and vanishes for $\phi=\varphi_0\pm\pi/2$. For an $x$-polarized input at 937 nm and a nanofiber of radius $a=250\ \mathrm{nm}$, $\frac{|E_z|^2}{|E|^2}\left(r=a_+,\phi=0\right)\simeq 20\%$. As a consequence, the polarization of a single propagating beam will be elliptical everywhere except for $\phi=\varphi_0\pm\pi/2$. The ellipticity of the beam will be maximum for $\phi=\varphi_0,\ \varphi_0+\pi$ as is illustrated in Fig. \ref{fig:sgle_polar}, giving rise to significant vector shifts, which we describe in section \ref{sec:results}.

\begin{figure}[h!]%
\centering
\begin{minipage}[c]{0.3\textwidth}
\includegraphics[width=\textwidth]{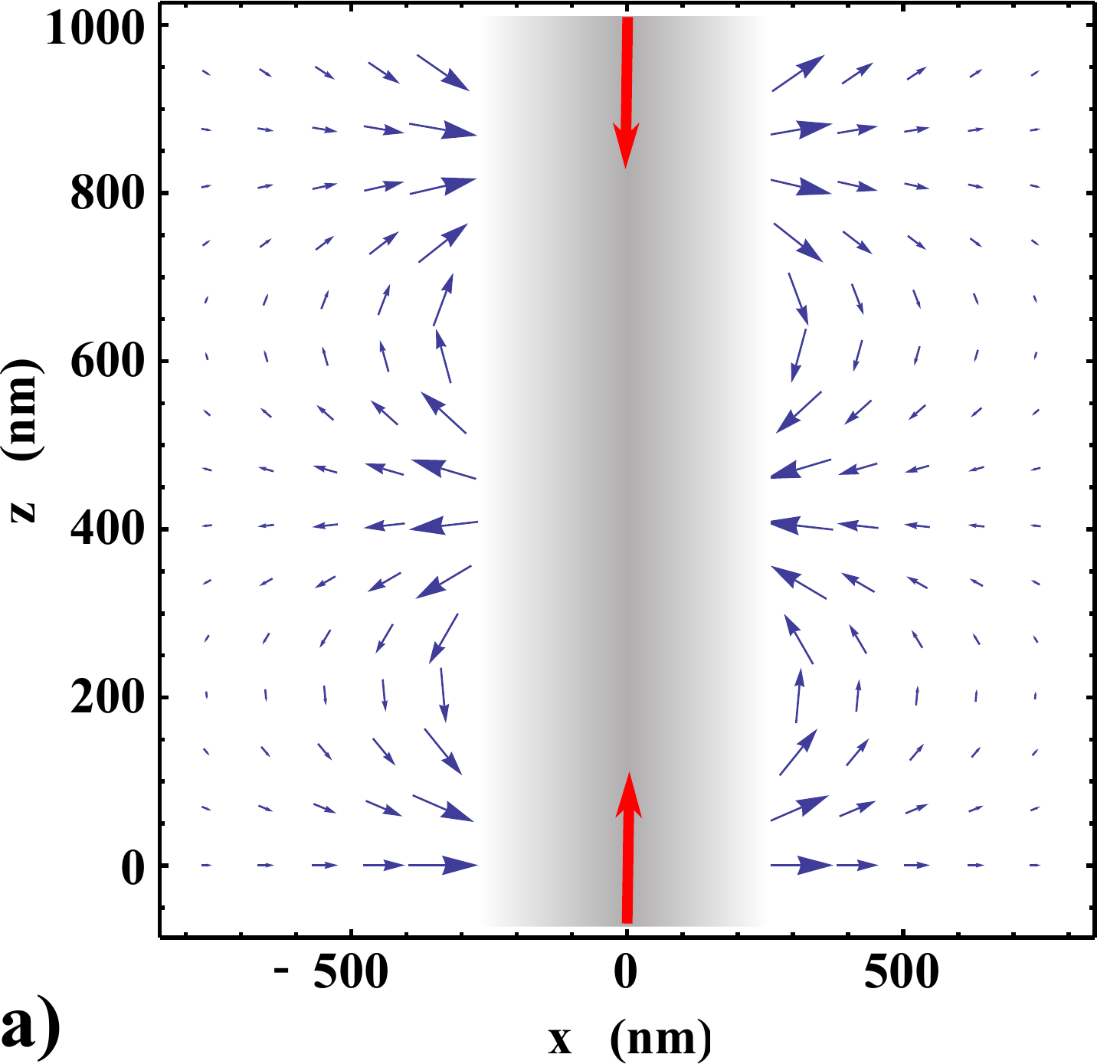}%
\end{minipage}
\hspace{0.02\textwidth}
\centering
\begin{minipage}[c]{0.3\textwidth}
\includegraphics[width=\textwidth]{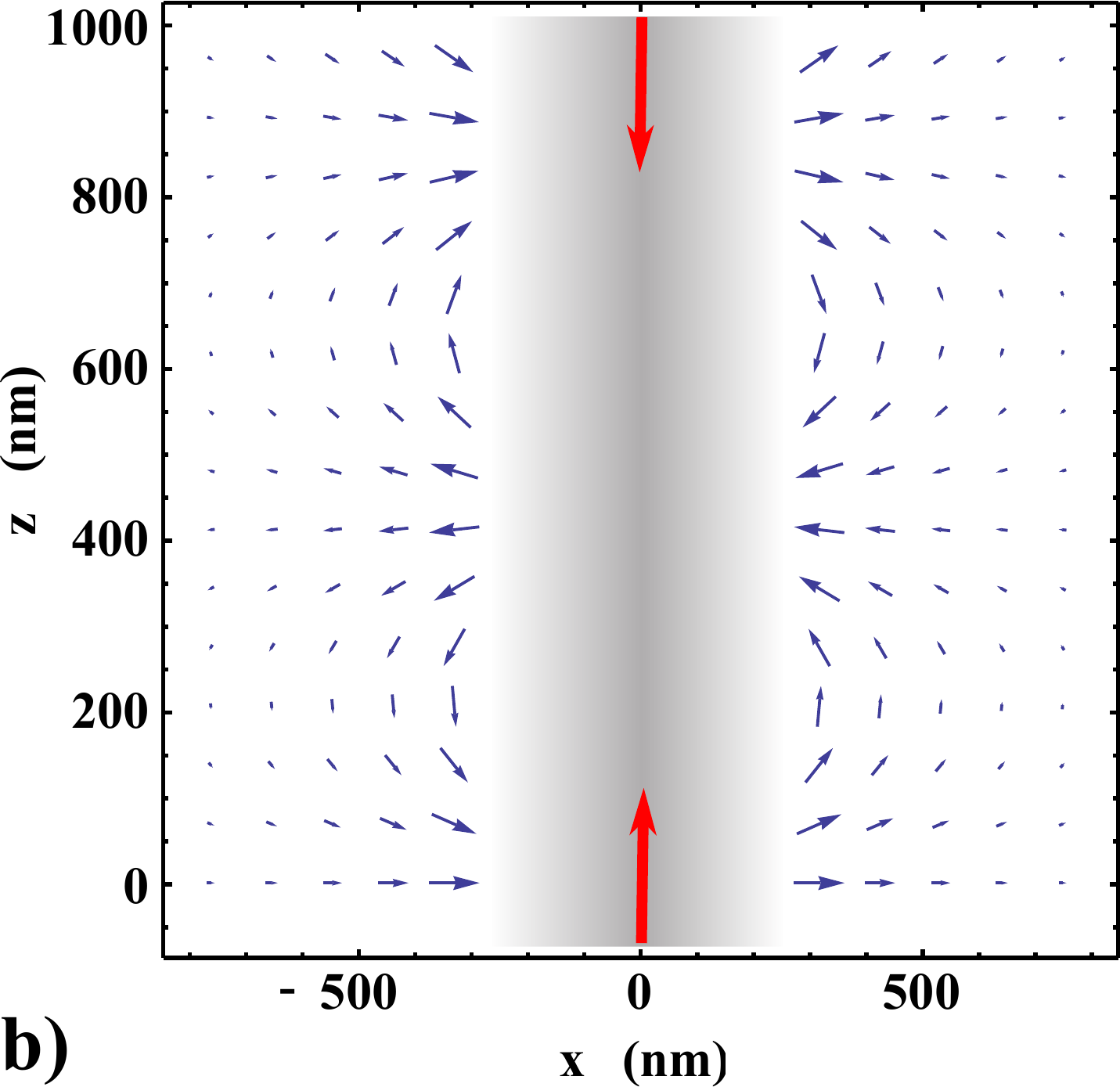}%
\end{minipage}
\hspace{0.02\textwidth}
\centering
\begin{minipage}[c]{0.3\textwidth}
\includegraphics[width=\textwidth]{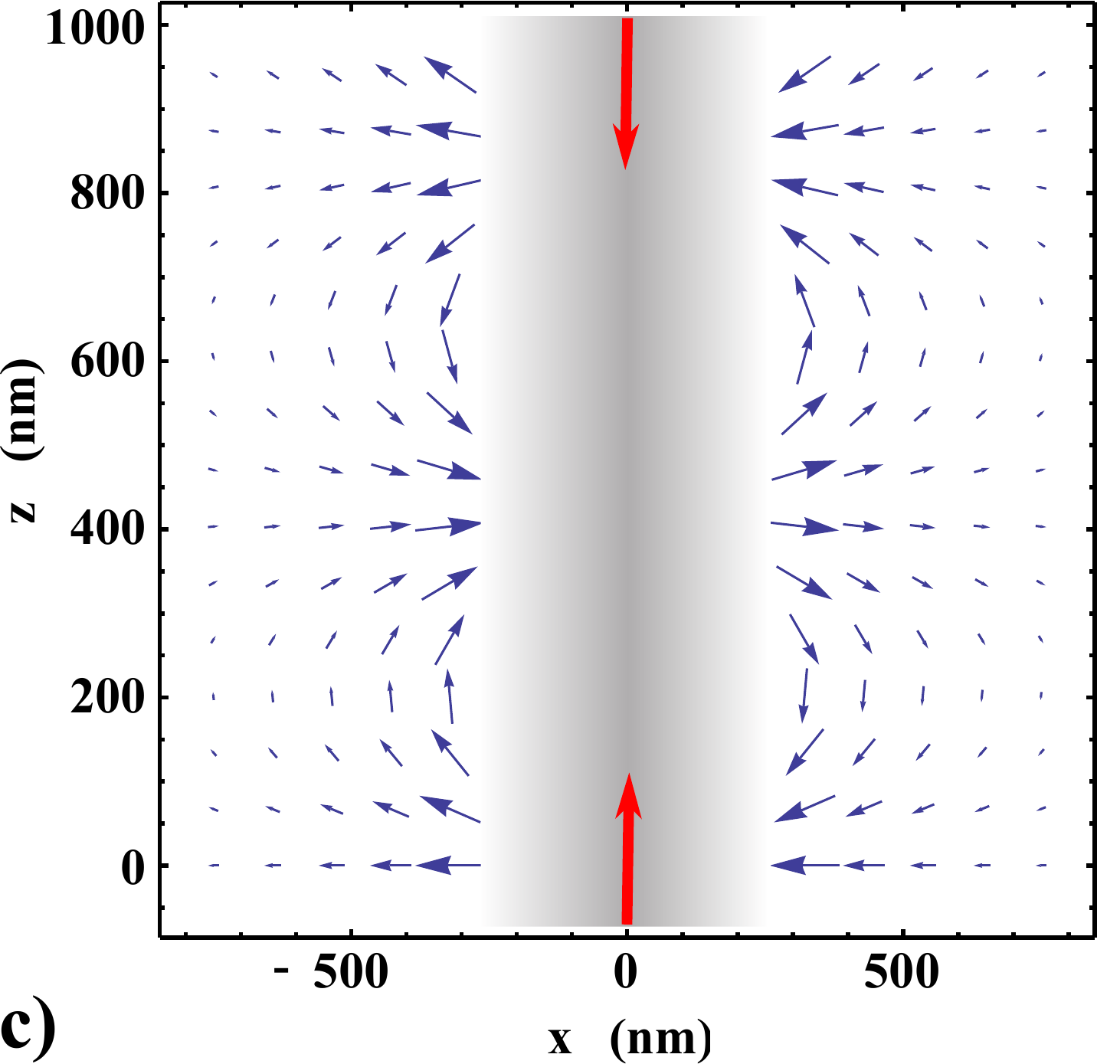}%
\end{minipage}
\caption{Total electric field $\mathbf{E}(x,y,z,t)$ for two counter-propagating beams in the plane $y=0$. The input beams are $x$-polarized. The electric field $\mathrm{Re}[\mathbf{E}(x,y,z,t)]$ is shown by the blue arrows. The red arrows indicate the beams propagation directions. The electric field is shown for \textbf{a)} $\omega t=0$, \textbf{b)} $\omega t=\pi/4$, and \textbf{c)} $\omega t=\pi$. As opposed to Fig. \ref{fig:sgle_polar}, the polarization of the electric field is linear at any point $|\mathbf{r}|>a$ (i.e., the polarization vector has no ellipticity and $\mathbf{E}$ does not rotate in time at a given position $\mathbf{r}$ as in \ref{fig:sgle_polar}).}%
\label{fig:comp_polar}%
\end{figure}

\begin{figure}[h!]
\centering
\begin{minipage}[c]{0.45\textwidth}
\includegraphics[width=\textwidth]{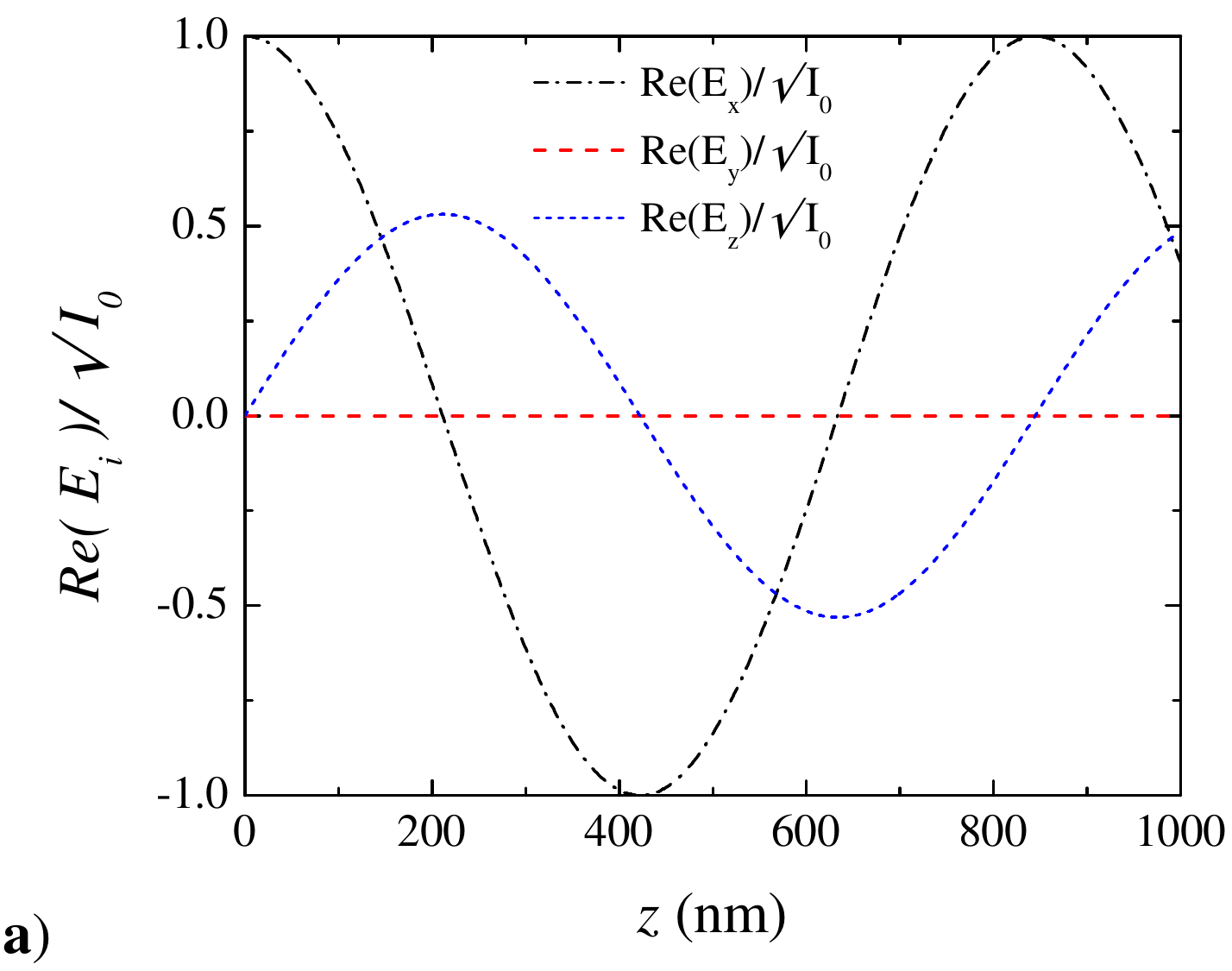}%
\end{minipage}
\hspace{.05\textwidth}
\centering
\begin{minipage}[c]{0.45\textwidth}
\includegraphics[width=\textwidth]{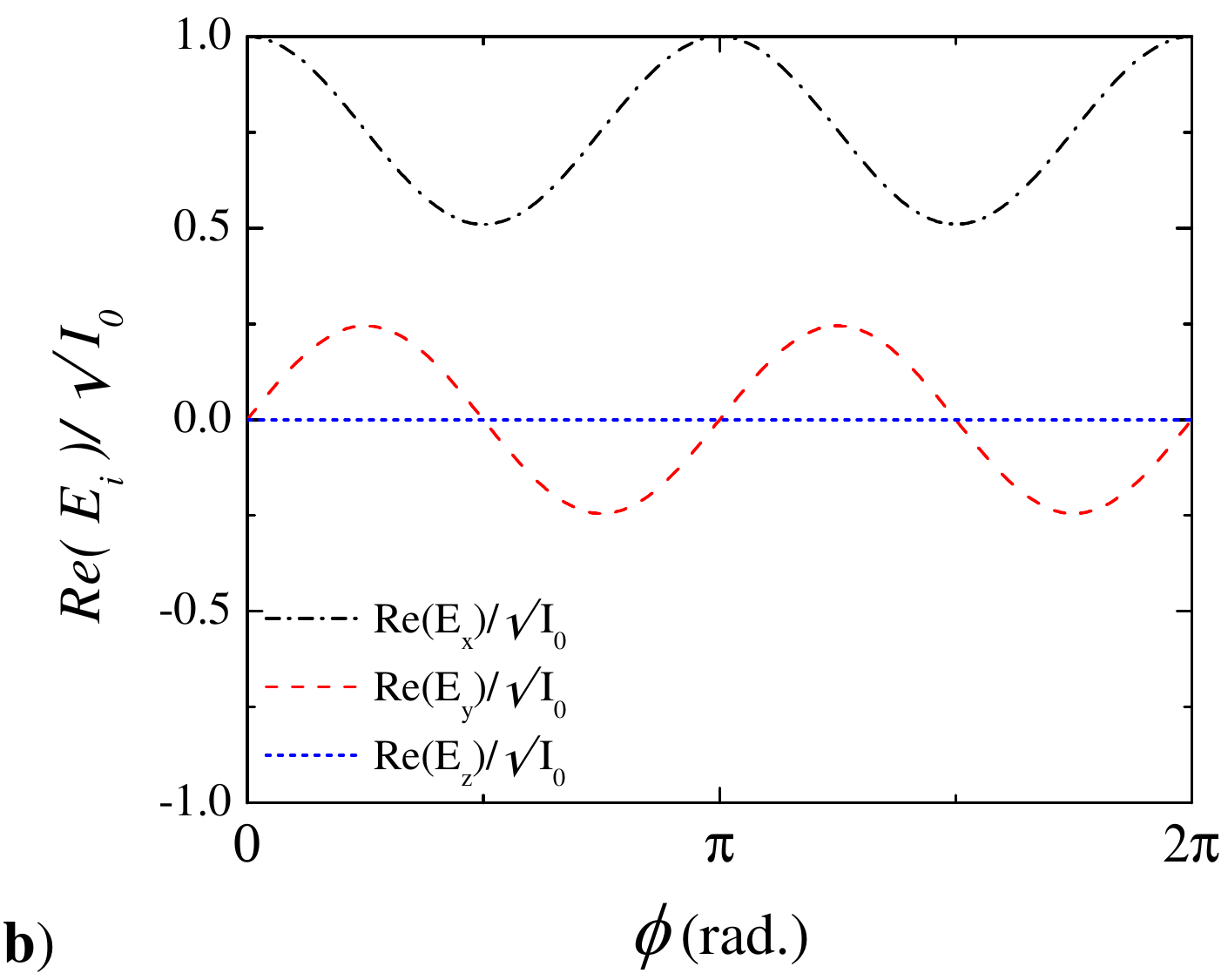}%
\end{minipage}
\caption{Electric field amplitude after interference $\mathbf{E}^{\mathrm{(tot)}}=\mathbf{E}^{\mathrm{(fwd)}}+\mathbf{E}^{\mathrm{(bwd)}}$ of two 937 nm beams with $\delta_{fb}=0$, at $t=0$ and $r=a_+$ as in Fig. \ref{fig:configs}d (i.e., $x$-polarized inputs with $\varphi_0=0$). The fields are normalized to the intensity $I_0$ at $r=a_+$, $\phi=0$, $z=0$. \textbf{a)} axial direction $z$ (at $\phi=0$). \textbf{b)}, azimuthal direction $\phi$ (at $z=0$). In particular, $\mathbf{E}^{\mathrm{(tot)}}$ has a fixed linear polarization at any given point $\mathbf{r}$ which rotates as $\mathbf{r}$ is varied.}
\label{fig:polar2}
\end{figure}

We can re-write Eqs. \eqref{eq:FundMode} as follows:

\begin{subequations}
\label{eq:FundMode2}
\begin{eqnarray}
E_{x}(r,\phi,z,t) & = & A e^{i(\omega t-\beta_{11}z)}, \\
E_{y}(r,\phi,z,t) & = & B e^{i(\omega t-\beta_{11}z)}, \\
E_{z}(r,\phi,z,t) & = & i C e^{i(\omega t-\beta_{11}z)},
\end{eqnarray}
\end{subequations}
where A, B, and C are real functions of $r$ and $\phi$. In particular, if one combines a forward-propagating beam $\mathbf{E}^{\mathrm{(fwd)}}$ expressed as Eq. \eqref{eq:FundMode2} with a backward-propagating beam of same amplitude and input polarization $\mathbf{E}^{\mathrm{(bwd)}}=A e^{i(\omega t+\beta_{11}z)}\mathbf{e}_x+B e^{i(\omega t+\beta_{11}z)}\mathbf{e}_y-i C e^{i(\omega t+\beta_{11}z)}\mathbf{e}_z$, the total field can be expressed as:
\begin{equation}
\mathbf{E^{\mathrm{(tot)}}}=\mathbf{E}^{\mathrm{(fwd)}}+\mathbf{E}^{\mathrm{(bwd)}}=2\left[A \cos(\beta_{11}z) \mathbf{e}_x+B \cos(\beta_{11}z) \mathbf{e}_y+C\sin(\beta_{11}z)\mathbf{e}_z\right]\cdot e^{i\omega t}.
\label{eq:lattice}
\end{equation}
The resulting electric field $\mathbf{E^{\mathrm{(tot)}}}=\mathbf{E}^{\mathrm{(fwd)}}+\mathbf{E}^{\mathrm{(bwd)}}$ forms an optical lattice with spatially rotating linear polarization as illustrated in Fig. \ref{fig:comp_polar}. In particular, the polarization state of the field rotates between the pure linear $x$ and $z$-polarizations along $z$ at $\phi=0$, as illustrated in Fig. \ref{fig:polar2}.

\subsection{Cancellation of the vector shifts} \label{sec:cancel}

In Ref. \cite{Vetsch2010}, the properties shown in Figs. \ref{fig:comp_polar} and \ref{fig:polar2} were used to cancel the vector shifts of ground and excited $m_F$ states in Cs for the pair of red-detuned trapping beams for the configuration in Fig. \ref{fig:configs}b. However, vector shifts due to the single blue-detuned beam were zero only for $\phi=\varphi_0, \ \varphi_0+\pi$. Although the atoms are trapped at $\phi=\varphi_0, \ \varphi_0+\pi$, inevitable fluctuations of the atom position will lead to non-zero vector light shifts of both ground and excited states.

The scheme in Fig. \ref{fig:configs}c allows for the use of reduced power for the blue-detuned beams as compared to Fig. \ref{fig:configs}b but with the consequence of large vector shifts from the ellipticity of the electric field even for $\phi=0$. We will therefore not consider this scheme in the next sections.

By contrast, the vector shifts of both the ground and excited states can be canceled for both the red and blue-detuned fields by using pairs of counter-propagating beams, as shown in Fig. \ref{fig:configs}d. In the $x-z$ plane, the vector shift for each pair becomes $\hat{H}_{\mathrm{1}}\propto(\alpha^{(1)}(\omega^{\mathrm{(fwd)}})-\alpha^{(1)}(\omega^{\mathrm{(bwd)}}))\frac{\hat{F_y}}{F}$ with $\omega^{\mathrm{(fwd)}}\simeq \omega^{\mathrm{(bwd)}}$, where $\omega^{\mathrm{(fwd,bwd)}}$ are the angular frequencies for the forward and backward propagating beams, and $\delta\pm\delta_{fb}/2$ are their detunings from the atomic transition frequency $\omega_a$, with two-photon detuning $\delta_{fb}=\omega^{\mathrm{(fwd)}}-\omega^{\mathrm{(bwd)}}$. For an atom in the $x-z$ plane, the total electric field is also contained in the $x-z$ plane, such that the scalar product $\left(\mathbf{\hat{E}^{(-)}}\times \mathbf{\hat{E}^{(+)}} \right) \cdot \mathbf{\hat{F}}$ in Eq. \eqref{eq:hamiltonian} is proportional to $\hat{F}_y$.

In the case of the red-detuned lattice, $\omega^{\mathrm{(fwd)}}_{\mathrm{red}}= \omega^{\mathrm{(bwd)}}_{\mathrm{red}}$ and $\hat{H}^{\mathrm{(red)}}_{\mathrm{1}}=0$, precisely as in Ref. \cite{Vetsch2010}. Adding a blue-detuned lattice with $\delta_{fb}=0$ would result in two superimposed lattices with unmatched spatial periods $2\pi/\beta^{\mathrm{red}}_{11}$, $2\pi/\beta^{\mathrm{blue}}_{11}$. To avoid this effect, the interference between the counter-propagating blue-detuned fields  $\mathbf{E}_{\mathrm{blue}}^{\mathrm{(fwd)}}$ and $\mathbf{E}_{\mathrm{blue}}^{\mathrm{(bwd)}}$ can be averaged over times short compared to the time scale of the motional and internal dynamics of a trapped atom by offsetting the frequencies of the two fields by  $\delta_{fb}\gg (\omega_{\mathrm{trap}},\delta_{\mathrm{hfs}})$, where $\omega_{\mathrm{trap}}$ and $\delta_{\mathrm{hfs}}$ are the trap angular frequency and the hyperfine splitting for the ground state, respectively. This will also suppress spurious two-photon processes (e.g., two-photon Stark shift \cite{Ashraf2001}) as well as parametric heating due to intensity modulation \cite{Savard1997}.

For $\omega^{\mathrm{(fwd,bwd)}}_{\mathrm{blue}}=\omega_a+(\delta\pm \delta_{fb}/2)$, we achieve a vector shift cancellation for the blue-detuned field to the first order in $1/\delta$, namely:
\begin{equation}
\hat{H}_{\mathrm{1}}^{\mathrm{(blue)}}\propto \frac{\delta_{fb}}{\delta^2}\frac{\hat{F_y}}{F} +\mathcal{O}(1/\delta^3).
\label{eq:correction}
\end{equation}
For typical values of $\delta=85 \ \mathrm{THz}$ and $\delta_{fb}=30 \ \mathrm{GHz}$, $\delta_{fb}/\delta=3.5\times10^{-4}$.

\subsection{Magic wavelengths for an evanescent field trap}

\begin{figure}[h!]\centering
\includegraphics[width=0.45\columnwidth]{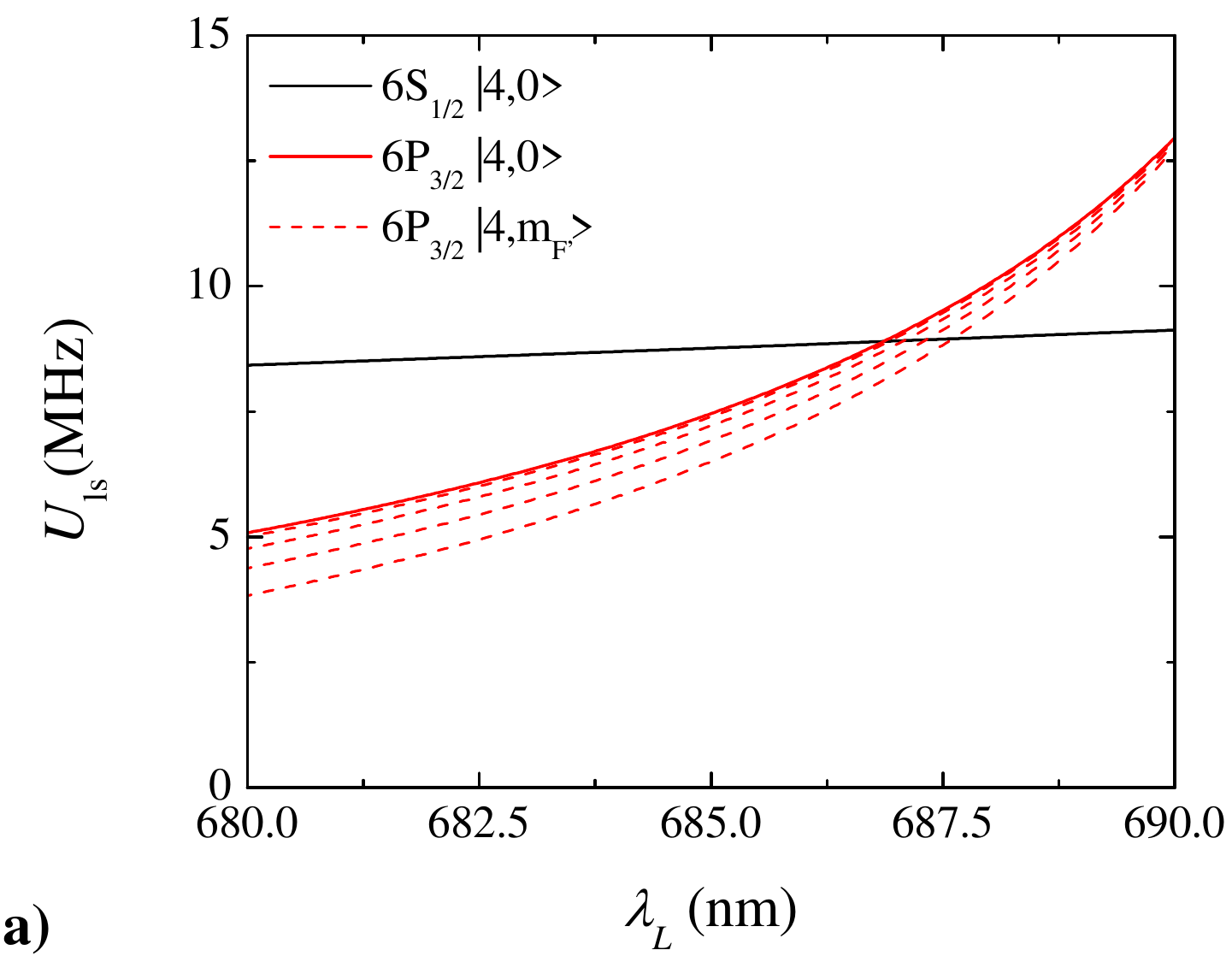}%
\includegraphics[width=0.45\columnwidth]{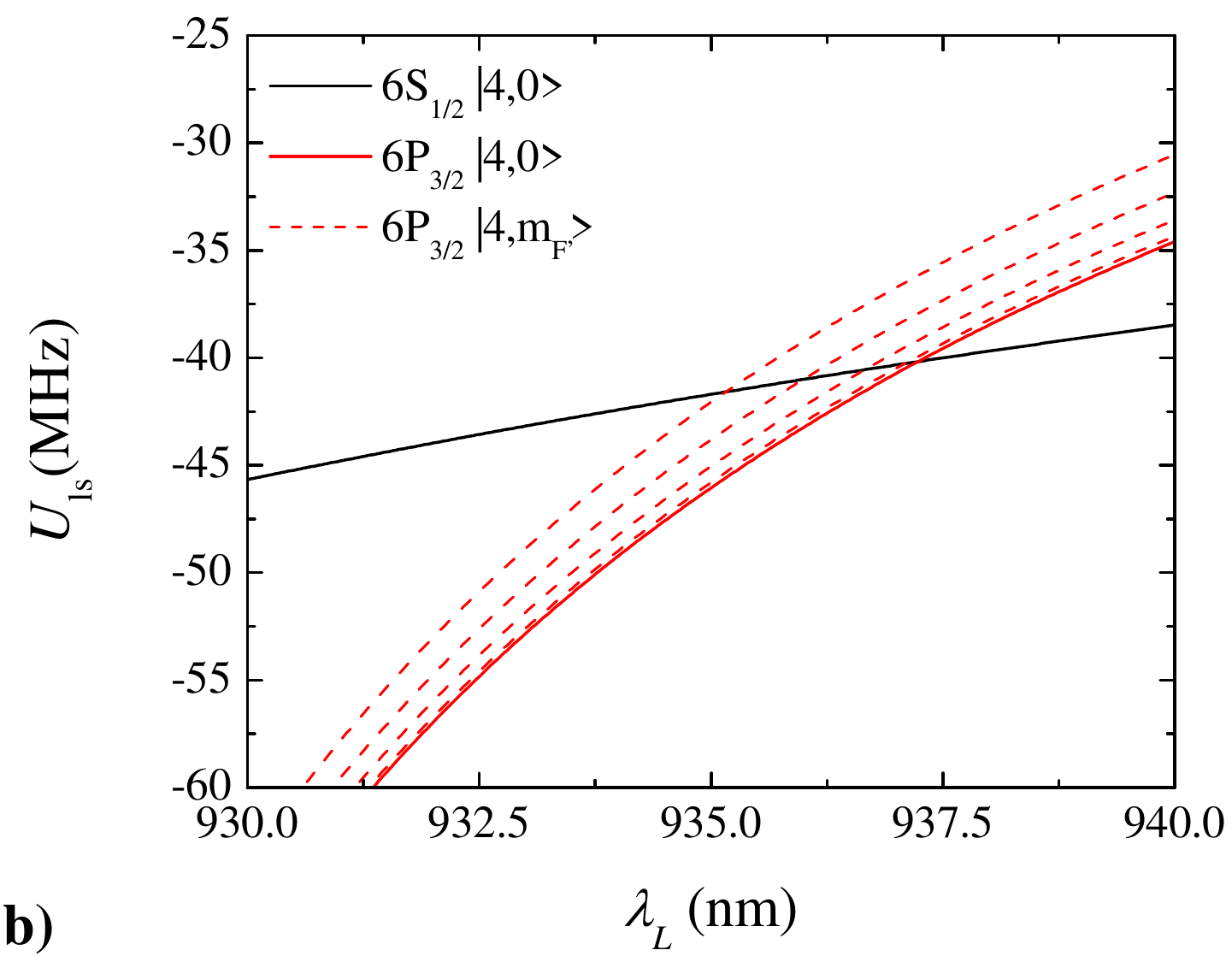}%
\caption[Searching for magic wavelengths of $D_2$ line of Cesium]{Magic wavelengths of the Cs $D_2$ line. We display the light shift $U_{\mathrm{ls}}$ for a linearly polarized beam with constant intensity $2.9\times10^9\mathrm{W/m^2}$ around \textbf{a)}, the blue-detuned magic wavelength at $\lambda_{\mathrm{blue}}\simeq687$ nm, and \textbf{b)}, red-detuned magic wavelength at $\lambda_{\mathrm{red}}\simeq937$ nm.}
\label{AppendixCmagicplot}
\end{figure}

To make the nanofiber trap state-insensitive, it is necessary to cancel the differential scalar shift $\delta U_{\mathrm{scalar}}$ by operating the trap at the magic wavelengths, as proposed in Ref. \cite{Kien2005a}, in which only the effects of the scalar and tensor shifts were considered. Here we deal with the full complexity of the vector field $\mathbf{E}(\mathbf{r})$ and the resulting vector light shifts. We numerically determine the red-detuned and blue-detuned magic wavelengths, following the procedure described in Refs. \cite{Ye2008,McKeever2003,Katori2003,Arora2007}. We include all the hyperfine levels $F$ and Zeeman sublevels $m_F$ of the electronic states $\{6S_{1/2},\cdots,15S_{1/2}\}$, $\{6P_{1/2},\cdots,11P_{1/2}\}$, $\{6P_{3/2},\cdots,11P_{3/2}\}$, $\{6D_{3/2},\cdots,11D_{3/2}\}$, and  $\{6D_{5/2},\cdots,11D_{5/2}\}$. The effect of the tensor shifts on the excited state is manifest in the quadratic splitting of the $m_{F'}$ sublevels (Fig. \ref{AppendixCmagicplot}). We find a red-detuned magic-wavelength located around 935nm, in accordance with the previously published values \cite{McKeever2003,Arora2007}. In the next sections, we will use the value $\lambda_{\mathrm{red}}=937\ \mathrm{nm}$, that cancels $\delta U_{\mathrm{scalar}}$ for the $6P_{3/2}$ excited state $|F'=4,m_{F'}=0\rangle$. We choose $F'=4$ due to its relevance to coherent two-photons processes \cite{ Boozer2007, Choi2008, Hammerer2010}. There are several blue-detuned magic wavelengths \cite{Kien2005a,Arora2007}. For our trap, we use the magic wavelength $\lambda_{\mathrm{blue}}$ at approximately 687 nm \cite{Kien2005a}. Since this is the second closest blue-detuned magic wavelength to 852 nm, it has the second highest ground-state polarizability and therefore requires the second lowest optical intensity to generate the required trapping potential (we do not consider the magic wavelength at $792 \ \mathrm{nm}$, as it is too close to the $8S_{1/2}$ to $6P_{3/2}$ transition at $794 \ \mathrm{nm}$).

We have neglected higher order processes in our analysis, including two-photon and electric quadrupole transitions, near 687 nm \cite{DiBerardino1998}.

\section{Numerical results: trapping potentials} \label{sec:results}

Using the atomic interaction Hamiltonian in the dipole approximation with the actual polarization profile of the evanescent field, we proceed to analyze the adiabatic potentials for the nanofiber trap for a Cs atom in its $6S_{1/2}$ ground and $6P_{3/2}$ excited states.

\subsection{Total potential} \label{sec:totalpot}

For a specific atomic state of Cs, the total atomic trap potential $U_{\mathrm{trap}}$ consists of the total light-shift potential $U_{\mathrm{ls}}$ calculated from the full Stark shift Hamiltonian (Eq. \eqref{eq:hamiltonian}), as well as the surface interaction potential of an atom with the dielectric waveguide $U_{\mathrm{surface}}$, namely

\begin{equation}
U_{\mathrm{trap}}=U_{\mathrm{ls}}+U_{\mathrm{surface}}.
\label{eq:trappot}
\end{equation}

The Casimir-Polder interaction between the atom and dielectric surface has a significant effect on the atomic motion at distance scales near 100 nm \cite{Alton2011,Sukenik1993,Bordag2001,Sandoghdar1992, Hinds1997}. The surface potential of a ground state Cs atom near a planar dielectric surface can be reasonably approximated by the van der Waals potential which decays as $d^{-3}$, where $d = r-a$:
\begin{equation}
U_{\mathrm{surface}}=- \frac{C_3}{d^3},
\label{eq:surfpot}
\end{equation}
where we use $C_3/h=1.2\ \mathrm{kHz \ \mu m^3}$ \cite{Stern2011}.  Because the retarded Casimir-Polder forces decrease faster away from the surface than the van der Waals forces, using $U_{\mathrm{surface}}$ overestimates the surface interaction at the trap location $d\approx\ 200 \ \mathrm{nm}$. Additionally, the curvature of the nanofiber cylindrical geometry reduces the potential strength even further \cite{Stern2011,Emig2006}. The $d^{-3}$ scaling of the van der Waals expression for a planar surface is therefore an overestimate of the actual surface potential, and is used for simplicity in the calculations presented here, with more complete expressions for Cs presented in Ref. \cite{Stern2011}. Furthermore, we neglect any dependence on the $m_{F'}$ sublevels of the excited state $6P_{3/2}$, and simply approximate $U_{\mathrm{surface}}^{6P_{3/2}}\approx2\ U_{\mathrm{surface}}$ \cite{Laliotis2007}.

We calculate the adiabatic potential of Eq. \eqref{eq:trappot} by diagonalizing the total interaction Hamiltonian $\hat{H}=\hat{H}_{\mathrm{ls}}+\hat{H}_{\mathrm{surface}}$ at each point in space, where $\hat{H}_{\mathrm{surface}}$ is the scalar surface Hamiltonian. At each point $\mathbf{r}(r, \phi,z)$, we obtain a set of eigenstates and the corresponding eigenenergies. These eigenstates are superpositions of the $|F,m_F\rangle$ bare Zeeman sublevels.  Due to the complex polarization of the trapping fields, the energy eigenstates are not necessarily eigenstates of any projection of the angular momentum operator.

\subsection{Effect of the light shifts in a ``non-magic'' trap}

First, we consider the trapping parameters for the experiment by Vetsch \textit{et al.} \cite{Vetsch2010}. Despite its impressive experimental success, the ground-state levels exhibit splittings that impair the ground state coherence. In the realization of Ref. \cite{Vetsch2010}, the two-color evanescent trap is constructed using a pair of counter-propagating $x$-polarized ($\varphi_0=0$) red-detuned beams $\mathbf{E}_{\mathrm{red}}=\mathbf{E}_{\mathrm{red}}^{\mathrm{(fwd)}}+\mathbf{E}_{\mathrm{red}}^{\mathrm{(bwd)}}$ ($P_\mathrm{red}=2\times2.2$ mW) at $\lambda_{\mathrm{red}}=1064$ nm, forming an optical lattice, and a single repulsive $y$-polarized ($\varphi_0=\pi/2$) blue-detuned beam $\mathbf{E}_{\mathrm{blue}}$ ($P_{\mathrm{blue}}=25$ mW)  at $\lambda_{\mathrm{blue}}=780$ nm. The SiO$_2$ tapered optical fiber has a radius $a=250$ nm in the trapping region.

\begin{figure}[t!]\centering%
\includegraphics[width=0.45 \columnwidth]{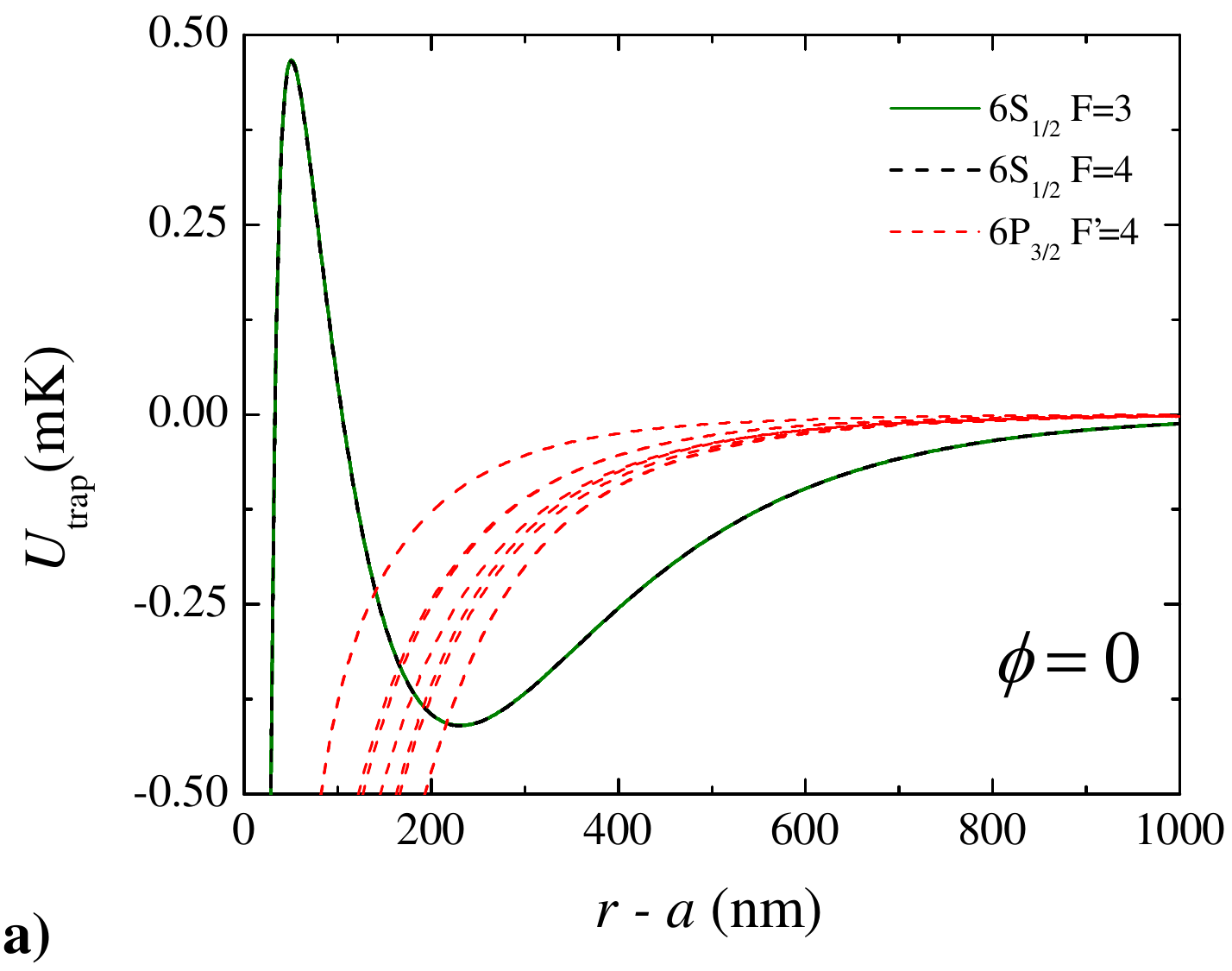}%
\includegraphics[width=0.45 \columnwidth]{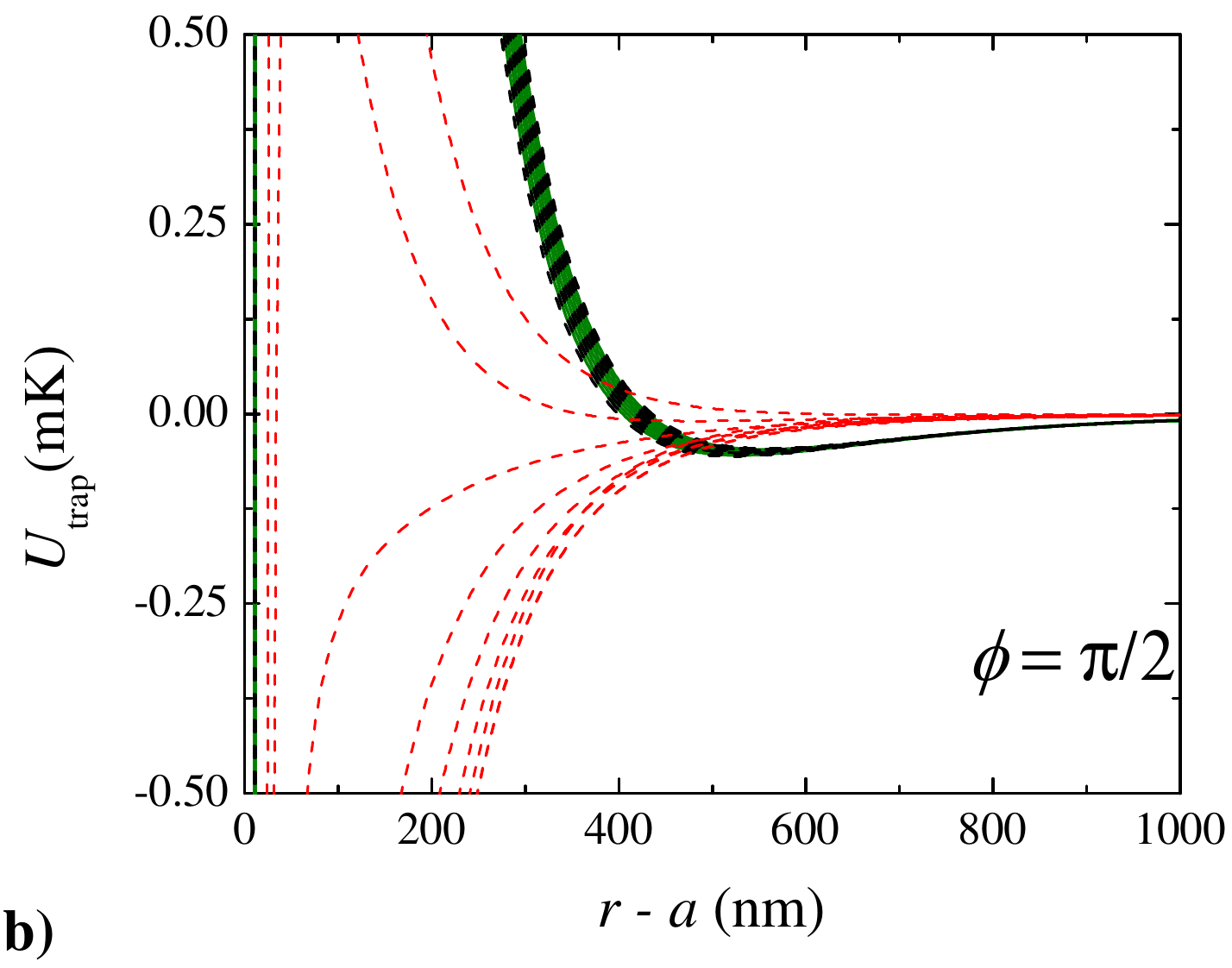}%
\caption{Radial dependence of the trapping potential of the ground and excited states for the parameters used in Ref. \cite{Vetsch2010} at $z=0$. The polarization configuration is the same as Fig. \ref{fig:configs}b. The energy sublevels of the ground states $F=3$ and $F=4$ of $6S_{1/2}$ are shown as  solid green and dashed black curves, and the $F'=4$ sublevels of the electronically excited state ($6P_{3/2}$) are shown as red dashed curves. \textbf{a)} Radial potential along $\phi=0$. The trap minimum is located about 230 nm from the fiber surface. The excited state is un-trapped, and split by the tensor shifts. \textbf{b)} Radial potential along $\phi=\pi/2$. Both ground and excited states are not trapped. The ground states exhibit a splitting due to the vector shifts induced by the elliptical polarization of the blue-detuned light, and the excited states are shifted by the vector and tensor shifts.}%
\label{fig:Vetsch_radial}%
\end{figure}

Fig. \ref{fig:Vetsch_radial} shows the radial trapping potential $U_{\mathrm{trap}}(r,\phi,z)$ of the ground states $F=3$ and $F=4$ of $6S_{1/2}$ and excited states $F'=4$ of $6P_{3/2}$, for $z=0$ and $\phi=0$ ($x$-axis) (Fig. \ref{fig:Vetsch_radial}a) and for  $\phi=\pi/2$ ($y$-axis) (Fig. \ref{fig:Vetsch_radial}b). The energy sublevels of the ground states at the trap location ($\phi=0$) are degenerate, as both trapping fields are linearly polarized as illustrated in Fig. \ref{fig:comp_polar}. The excited state energy sublevels are shifted due to the vector and tensor shifts. The trap depth for the ground state is $U_{\mathrm{depth}}=-0.4$ mK, located at $r-a\simeq230 \ \mathrm{nm}$ and $\phi=0$, whereas the excited states are not trapped at all.

\begin{figure}[t!]\centering%
\includegraphics[width=0.45 \columnwidth]{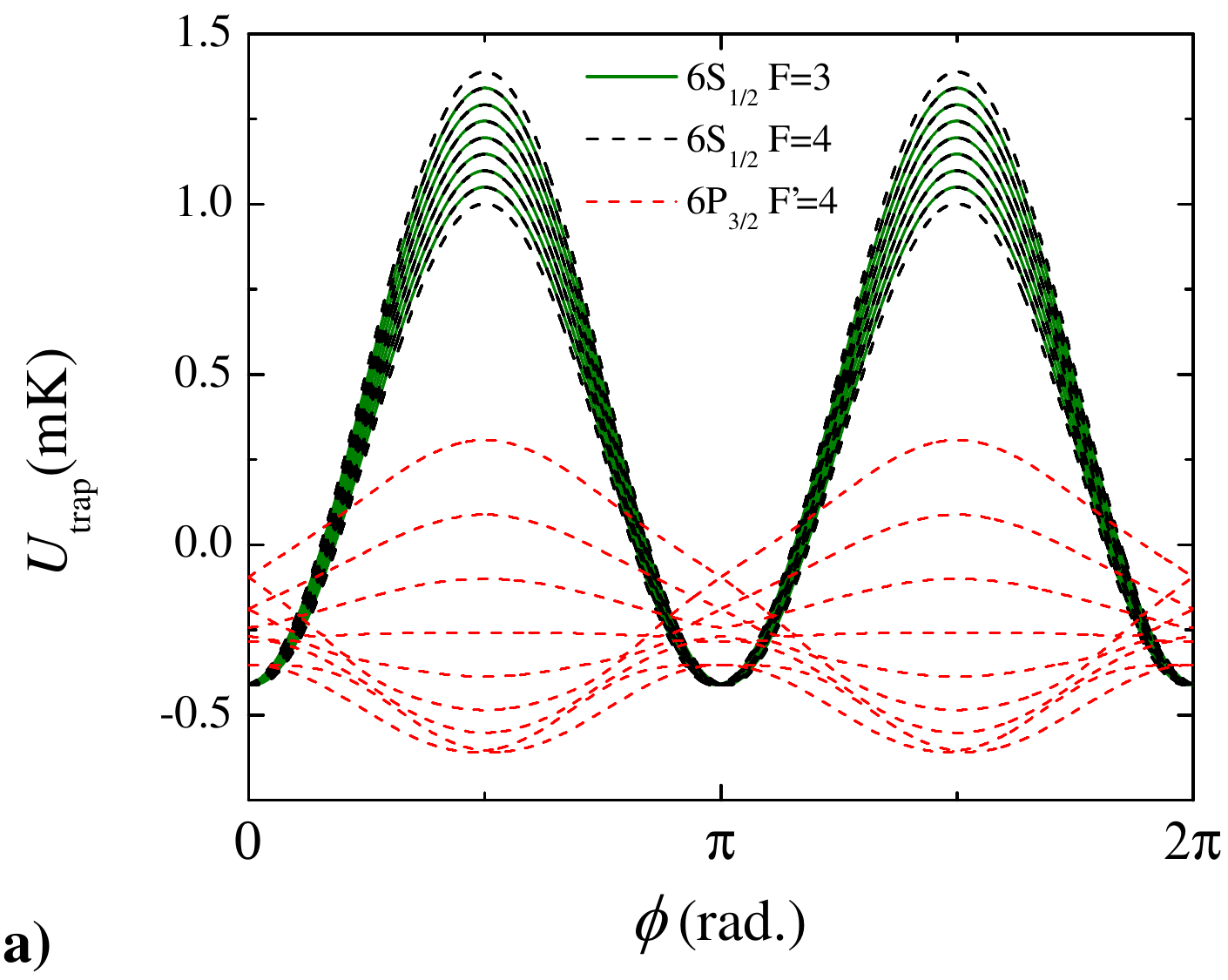}%
\includegraphics[width=0.45 \columnwidth]{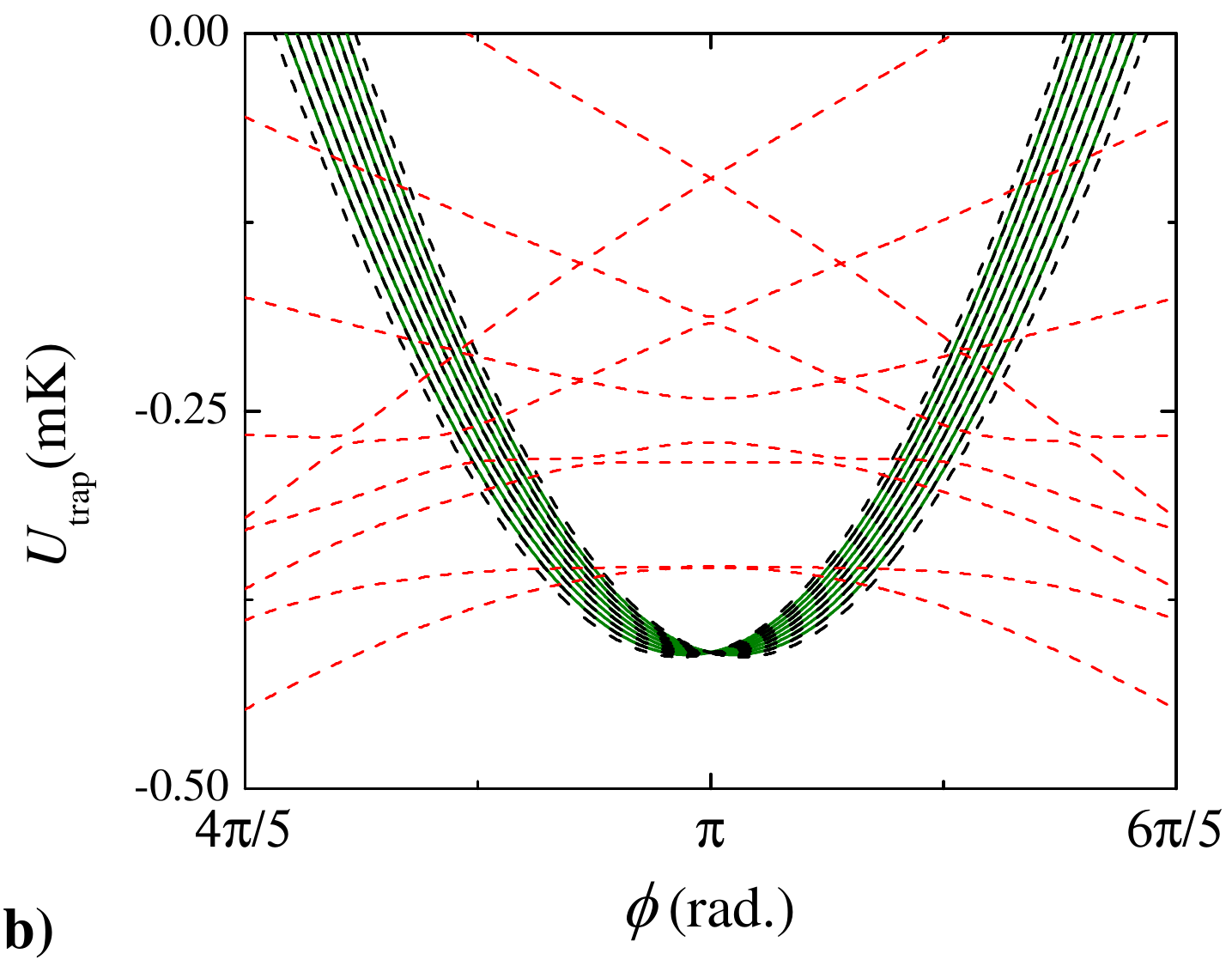}%
\caption{Azimuthal dependence of the trapping potential of the ground and excited states for the scheme in Fig. \ref{fig:configs}b and for the parameters used in Fig. \ref{fig:Vetsch_radial}. $r-a=230$ nm, $z=0$. \textbf{a)} The ground state splitting is minimum for $\phi=0$ and $\phi=\pi$. Everywhere else, the polarization of the blue-detuned field induces large vector shifts. \textbf{b)} Expanded view of \textbf{a)} near a trap minimum at $\phi=\pi$. The Zeeman-like splitting of the ground states is large even for small azimuthal angles. The excited-state level structures are greatly altered by the combined vector and tensor shifts.}%
\label{fig:Vetsch_azimuthal}%
\end{figure}

The azimuthal dependence of the trap potential reveals a significant inhomogeneous broadening of the energy sublevels due to the ellipticity of $\mathbf{E}_{\mathrm{blue}}$ for $\phi\neq 0,\pi$ (Fig. \ref{fig:Vetsch_azimuthal}). To estimate this broadening, we assume that the potential is harmonic around the trap minimum. By fitting the ground state $F=3$ potential with a harmonic potential around $\phi=\pi$, we obtain an azimuthal trapping frequency $\nu_{\mathrm{trap}_{\phi}}\simeq 150\ \mathrm{kHz}$. For an atom in its azimuthal motional ground state $|n\rangle_{\phi}=|0\rangle_{\phi}$ in such a potential, the half-width $\sigma_{r_{\phi}}$ of the corresponding single-atom distribution is given by $\sigma_{r_{\phi}} = \langle(r\phi)^2\rangle \simeq r_{\mathrm{trap}}\sigma_{\phi}= \sqrt{\frac{\hbar}{4\pi m\nu_{\mathrm{trap}_{\phi}}}}\simeq 16$ nm (or azimuthal half-width of $\sigma_{\phi}\simeq 2^{\circ}$). This leads to fast decoherence of the hyperfine and Zeeman levels, even with ground state cooling. Specifically, we estimate a spin-wave coherence time $\tau_m=1/\delta\nu_{\phi}\lesssim 5 \ \mu\mathrm{s}$, derived from the $\delta\nu_{\phi}=200\ \mathrm{kHz}$ splitting between the sublevels of the $F=4$ atomic ground state $16 \ \mathrm{nm}$ away from the trap minimum. This is significantly limited compared to the quantum memory performances of atomic ensembles in optical lattices (see, e.g., \cite{Radnaev2010}).

\begin{figure}[t!]\centering%
\includegraphics[width=0.45 \columnwidth]{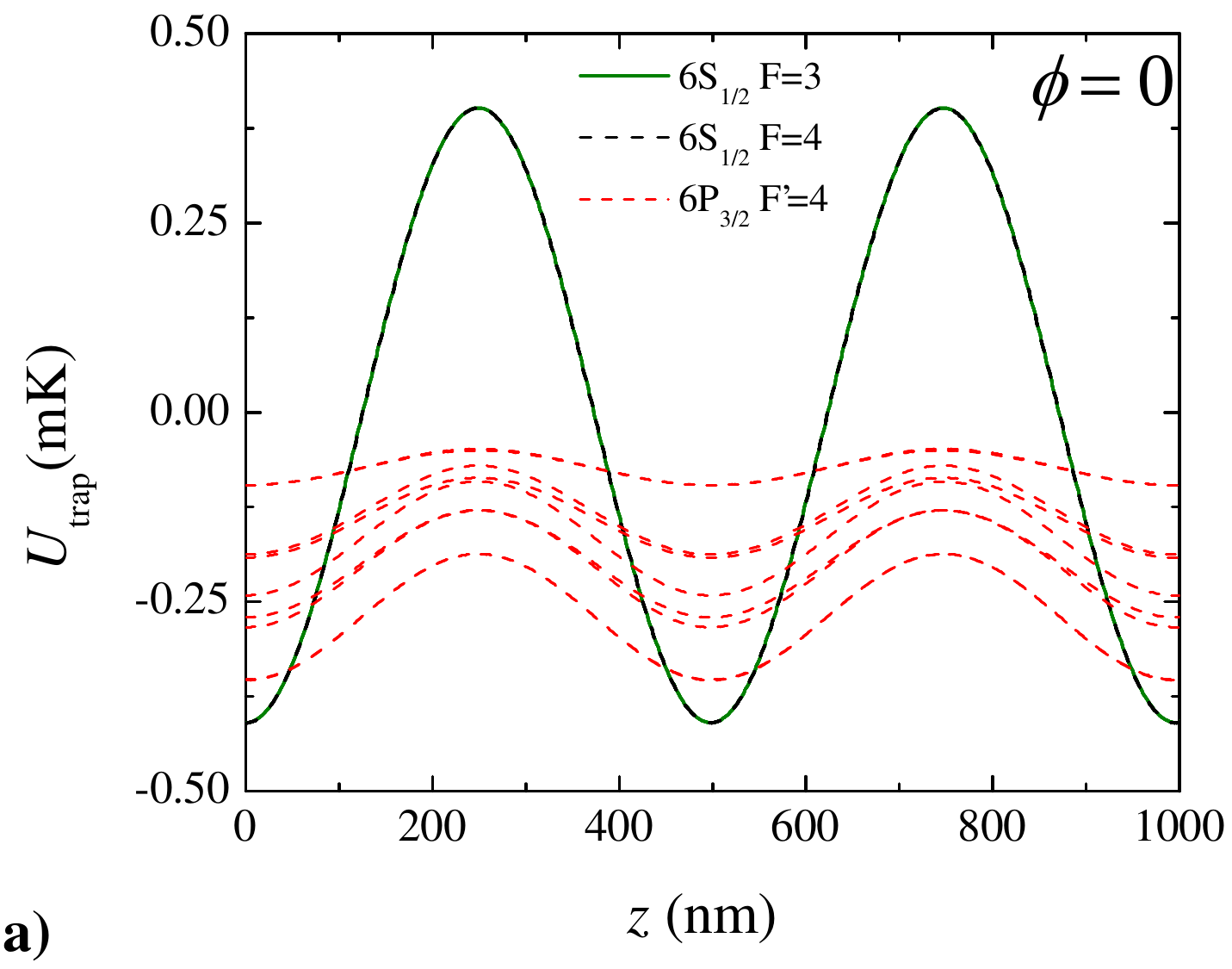}%
\includegraphics[width=0.45 \columnwidth]{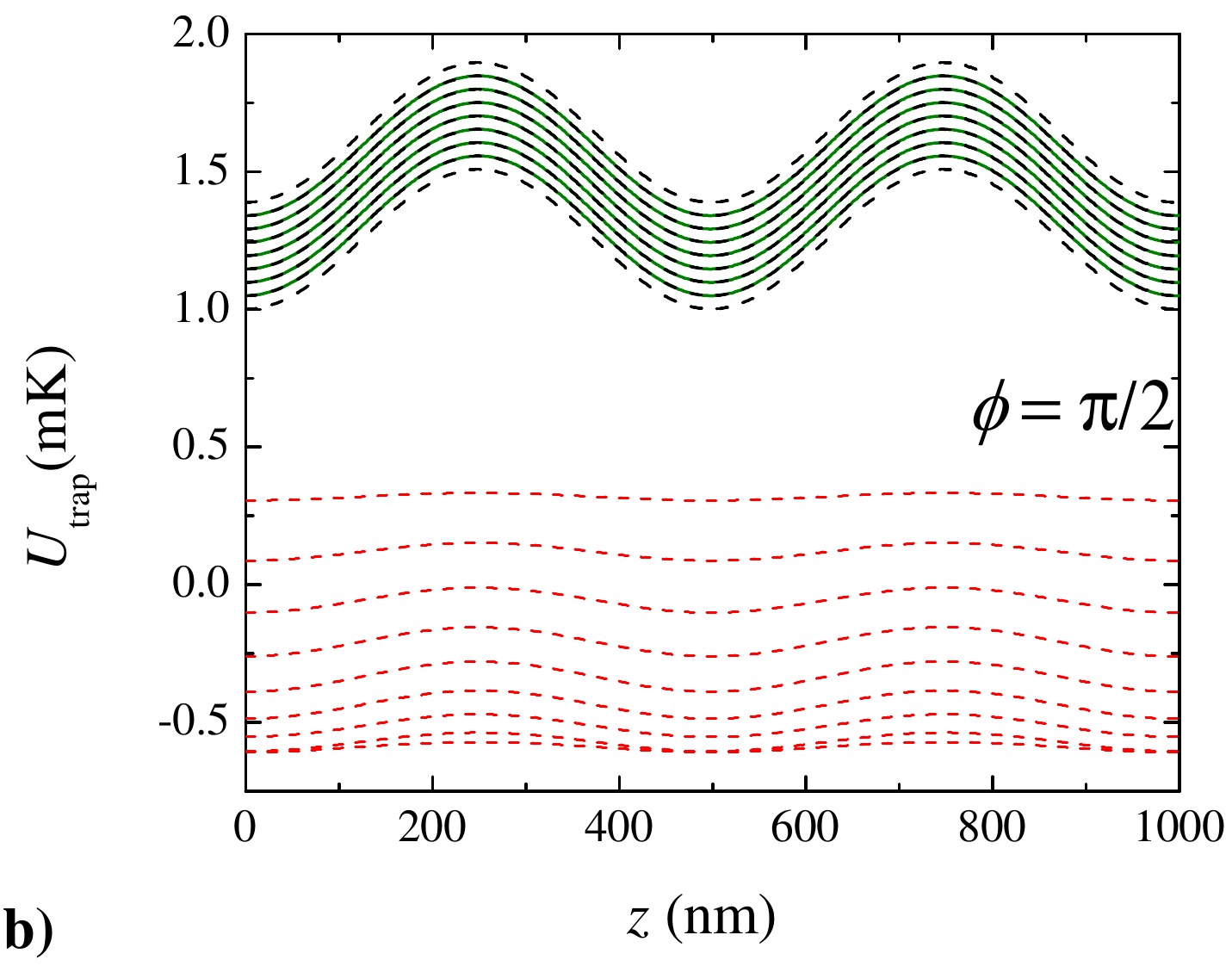}%
\caption{Axial dependence of the trapping potential for the ground and excited states for the scheme in Fig. \ref{fig:configs}b and for the parameters used in Fig. \ref{fig:Vetsch_radial}. \textbf{a)} Longitudinal potential along $\phi=0$. The distance from the fiber surface is set to $r-a=230\ \mathrm{nm}$ at the trap minimum. \textbf{b)} Longitudinal potential along $\phi=\pi/2$. The distance from the fiber surface is again set to $230\ \mathrm{nm}$.}%
\label{fig:Vetsch_axial}%
\end{figure}

Finally, we plot in Fig. \ref{fig:Vetsch_axial} a cross-section of the axial potentials showing the axial confinement of the ground and excited states.

The excited states are untrapped in all directions except along the fiber axis $z$ for the parameters of Ref. \cite{Vetsch2010}. An atom excited to these untrapped potentials will experience dipole-force fluctuations, leading to heating \cite{Corwin1999} and preventing near-resonant driving of the optical transition \cite{Vetsch2010a}.

\subsection{State-insensitive trapping potential}

Now, we analyze our proposed ``magic compensation'' scheme (as illustrated in Fig. \ref{fig:configs}d), demonstrating how using magic wavelength beams and compensating the trap ellipticity can reduce inhomogeneous broadening of the Zeeman sublevels in a nanofiber trap. For this trap, we use a pair of counter-propagating $x$-polarized ($\varphi_0=0$) red-detuned beams $\mathbf{E}_\mathrm{red}=\mathbf{E}_\mathrm{red}^{\mathrm{(fwd)}}+\mathbf{E}_\mathrm{red}^{\mathrm{(bwd)}}$ ($P_\mathrm{red}=2\times0.95 \ \mathrm{mW}$) at the ``magic'' wavelength $\lambda_\mathrm{red}=937 \ \mathrm{nm}$, forming a 1-D optical lattice. Counter-propagating, $x$-polarized blue-detuned beams at the second ``magic'' wavelength $\lambda_\mathrm{blue}=687 \ \mathrm{nm}$ are used with a power $P_\mathrm{blue}=2\times16 \ \mathrm{mW}$. The  resulting interference is averaged out by detuning the beams by $\delta_{fb}=30\ \mathrm{GHz}$, as explained in section \ref{sec:cancel}, leading to a first-order cancellation of vector light shifts as expressed by Eq. \eqref{eq:correction}. The beam intensities are chosen to generate a trap of similar depth as the one demonstrated in Ref. \cite{Vetsch2010}. The resulting adiabatic potential $U_{\mathrm{trap}}$ allows for state-insensitive 3D confinement of cold Cs atoms around a SiO$_2$ nanofiber of radius $a=250 \ \mathrm{nm}$.

\begin{figure}[h]
\centering%
\begin{minipage}[c]{0.45\textwidth}
\includegraphics[width=\textwidth]{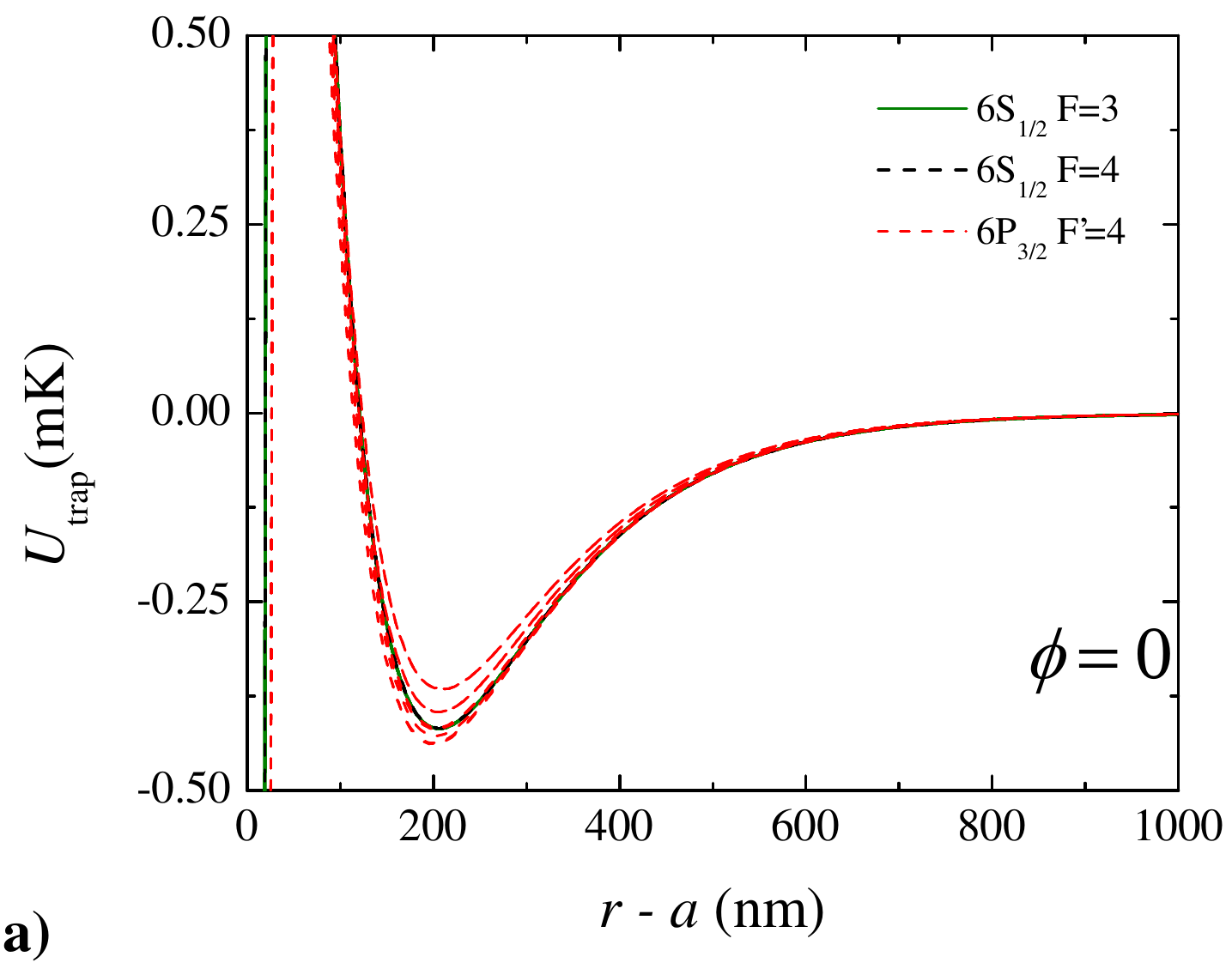}%
\end{minipage}
\centering
\begin{minipage}[c]{0.45\textwidth}
\includegraphics[width=\textwidth]{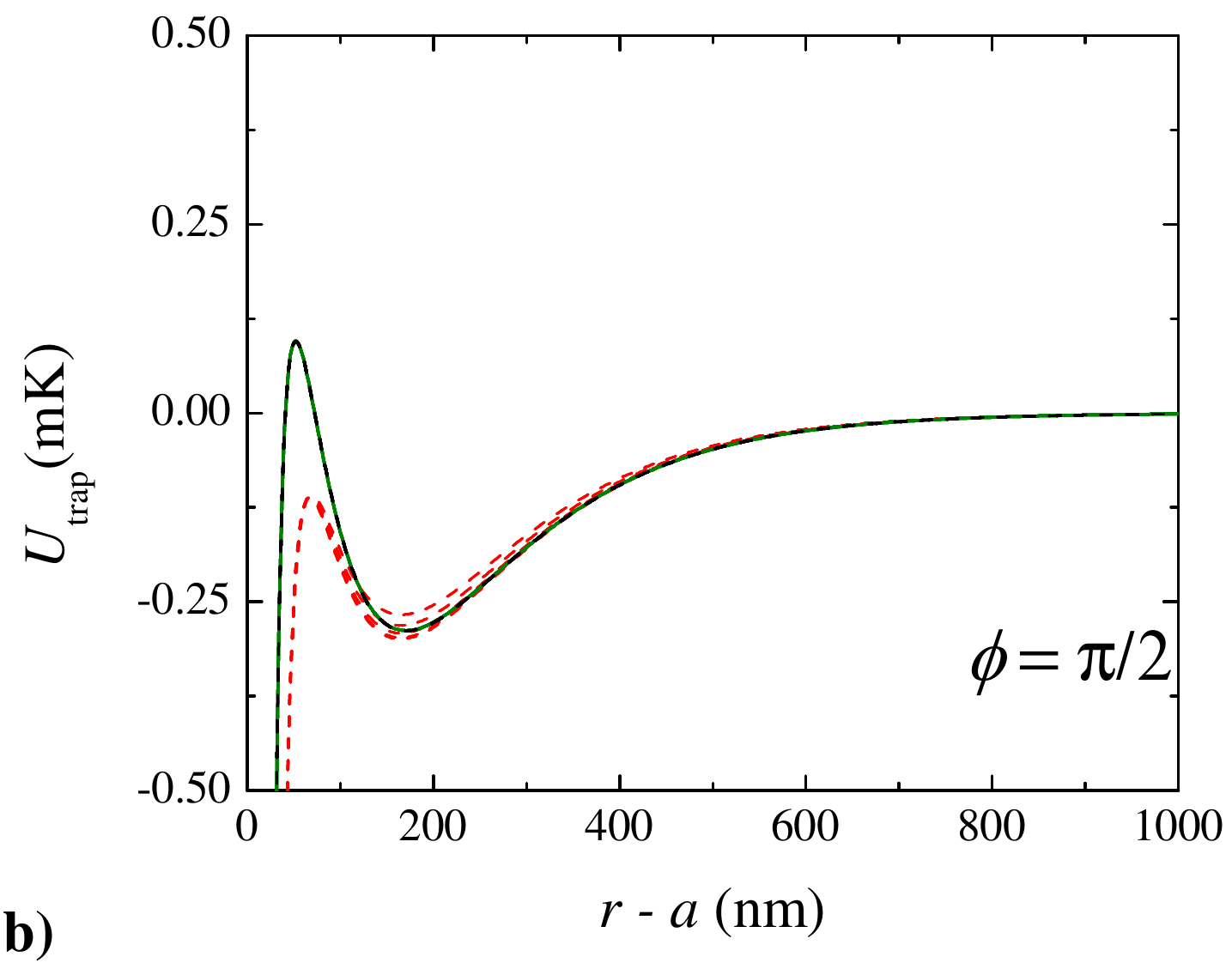}%
\end{minipage}
\centering
\begin{minipage}[c]{0.45\textwidth}
\includegraphics[width=\textwidth]{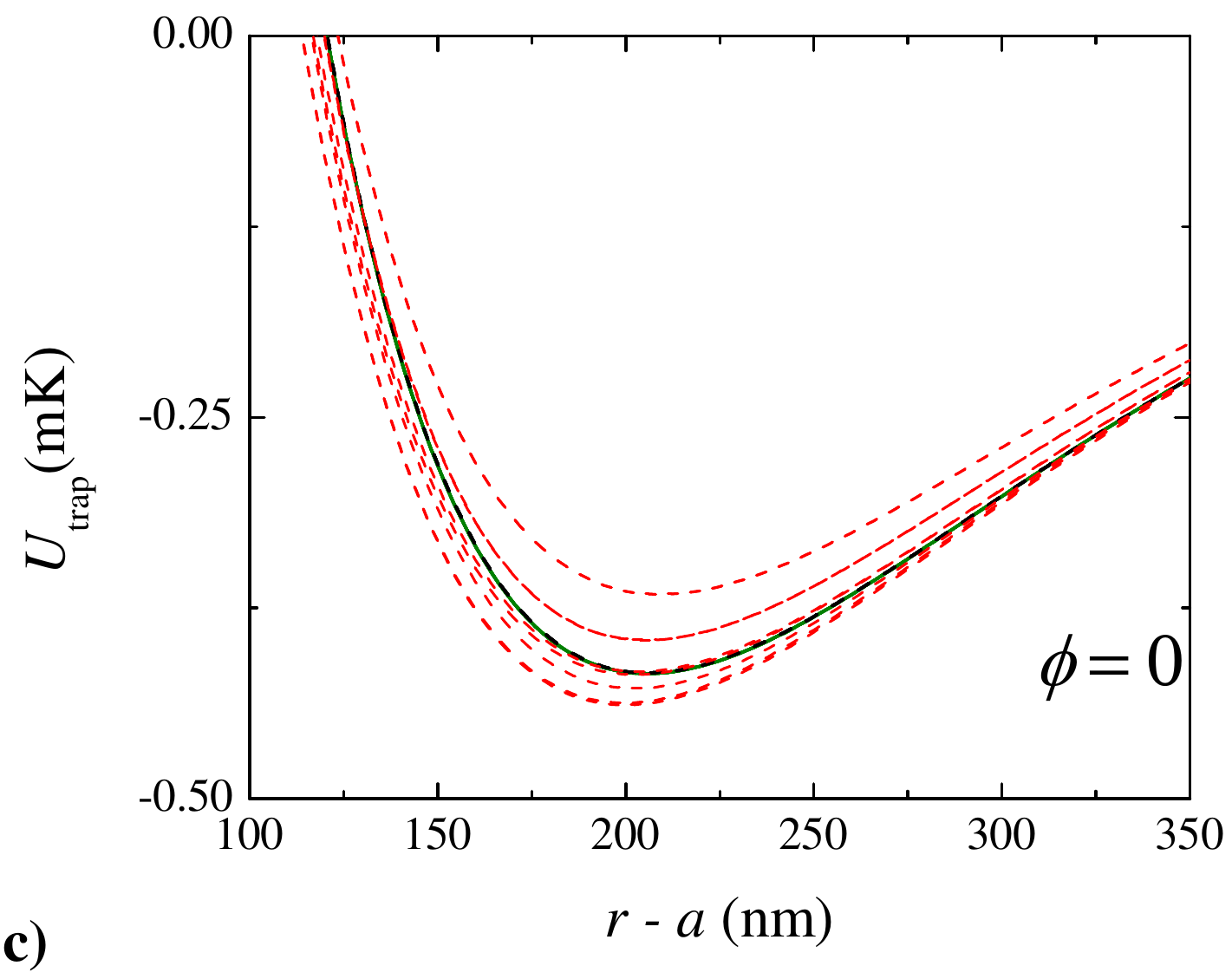}%
\end{minipage}
\centering
\begin{minipage}[c]{0.45\textwidth}
\includegraphics[width=\textwidth]{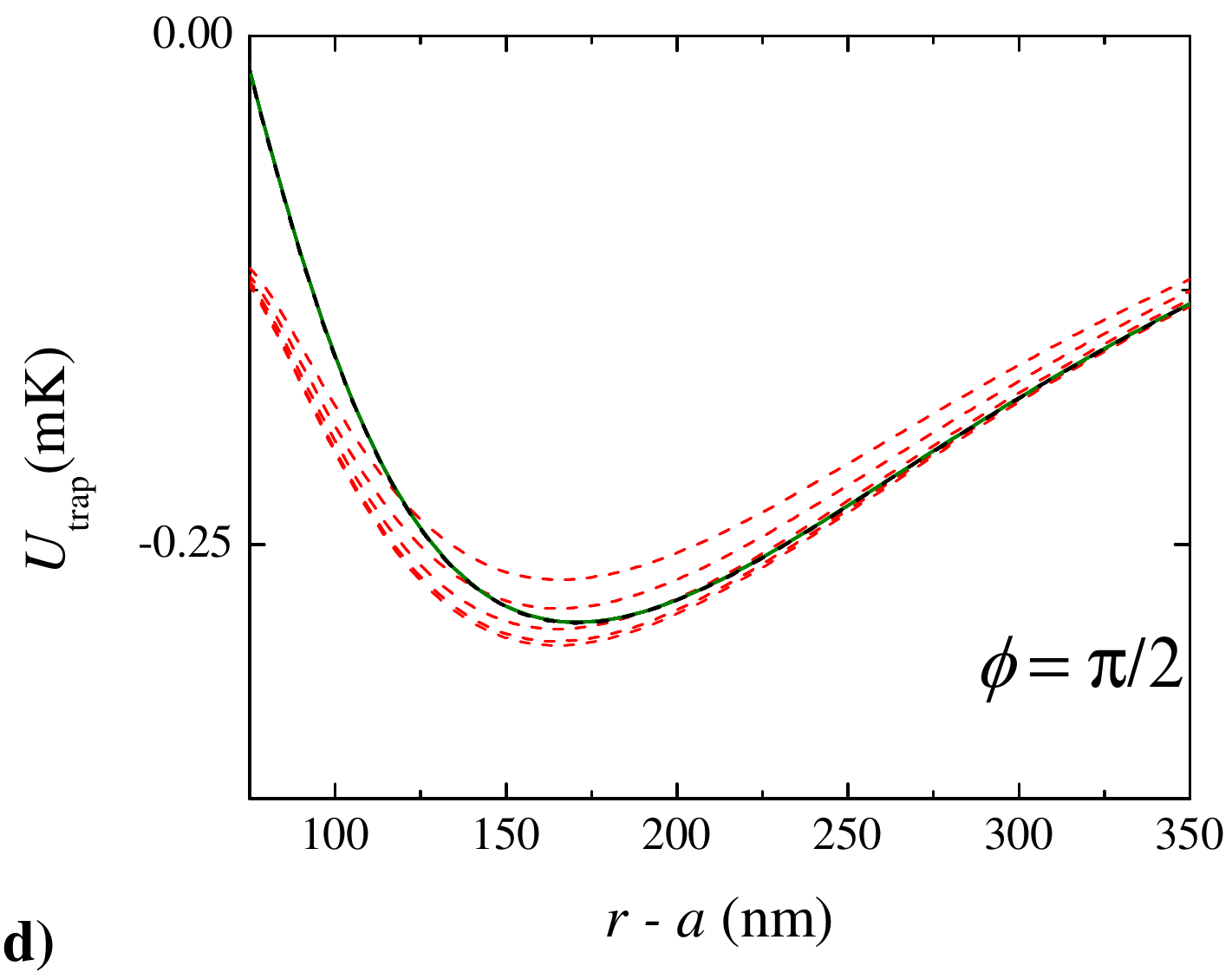}%
\end{minipage}
\caption{Radial dependence of the trapping potentials of the ground and excited states using the magic wavelengths and the configuration shown in Fig. \ref{fig:configs}d. All beams are polarized along $\phi=0$ (i.e., $\varphi_0$=0). The 937 nm beams each have a power of 0.95 mW. The 687 nm beams each have a power of 16 mW. \textbf{a)} Radial potentials along $\phi=0$ (i.e., $\varphi_0=0$). The trap minimum for $6S_{1/2}$ is located about 210 nm from the fiber surface. Both electronic ground and excited states are trapped, with residual splittings of the excited states due to the tensor shifts. \textbf{b)} Radial potential along $\phi=\pi/2$. \textbf{c), d)} Expanded view of \textbf{a), b)} around the trap minimum.}%
\label{fig:magic_comp_radial}%
\end{figure}

In Fig. \ref{fig:magic_comp_radial}, we show the radial trapping potential $U_{\mathrm{trap}}(r,\phi,z)$ of the ground and excited states for $z=0$, $\phi=0$ ($x$-axis) (Fig. \ref{fig:magic_comp_radial}a) and for  $z=0$, $\phi=\pi/2$ ($y$-axis) (Fig. \ref{fig:magic_comp_radial}b). Because the trapping fields are now effectively linearly polarized, the ground states are degenerate at both $\phi=0$ and $\phi = \pi/2$.  In contrast to a non-magic wavelength trap, the excited states are trapped with gradients that closely map that of the ground states. The sublevels of $6P_{3/2}$ are still non-degenerate due to the tensor shifts.  For $P_\mathrm{red}$, $P_\mathrm{blue}$ specified above, we find that the trap depth is $U_{\mathrm{depth}}=-0.4$ mK, located at $r-a\simeq r_{\mathrm{trap}}-a=210$ nm and $\phi=0,\pi$.

\begin{figure}[h]\centering%
\includegraphics[width=0.45\columnwidth]{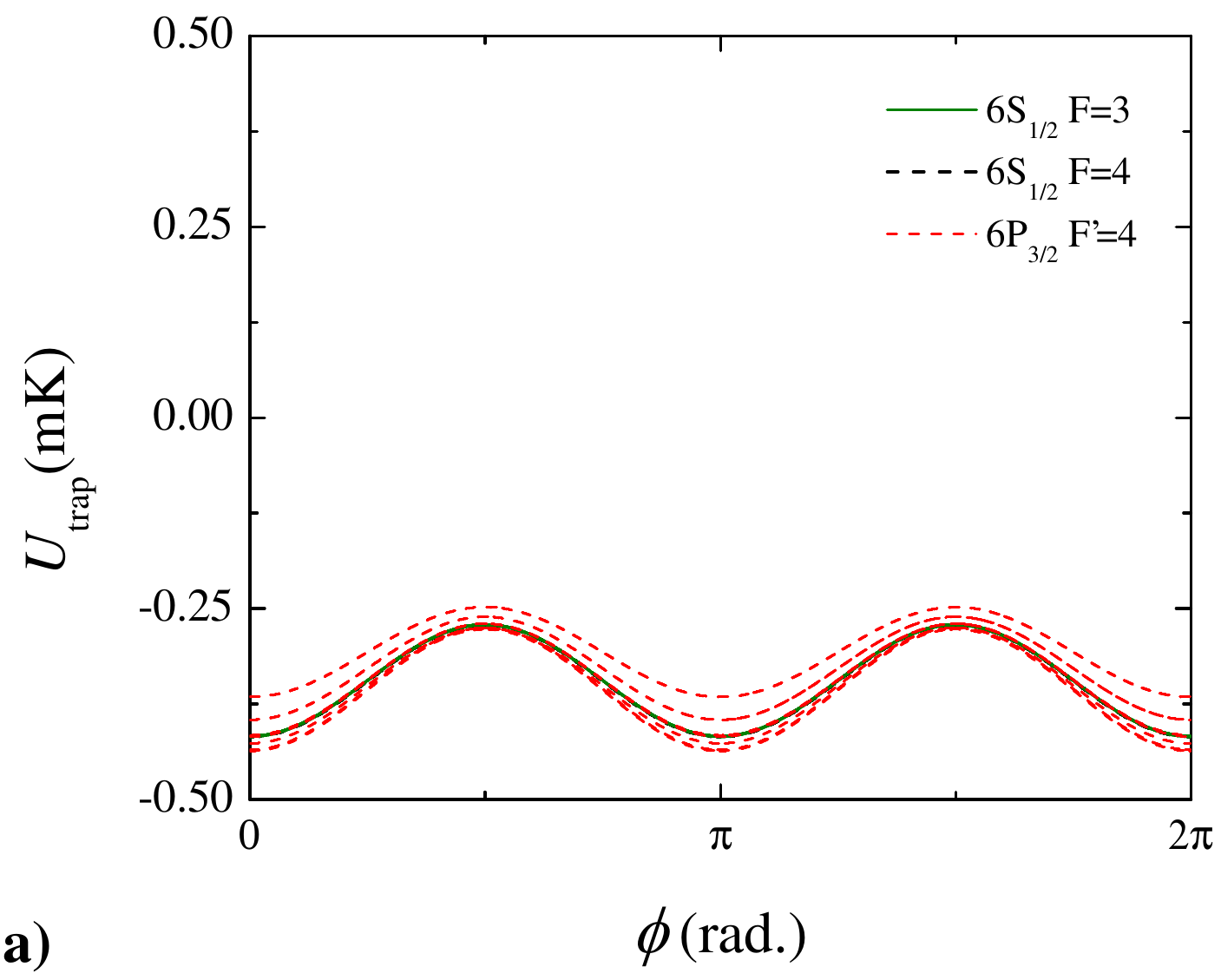}%
\includegraphics[width=0.45\columnwidth]{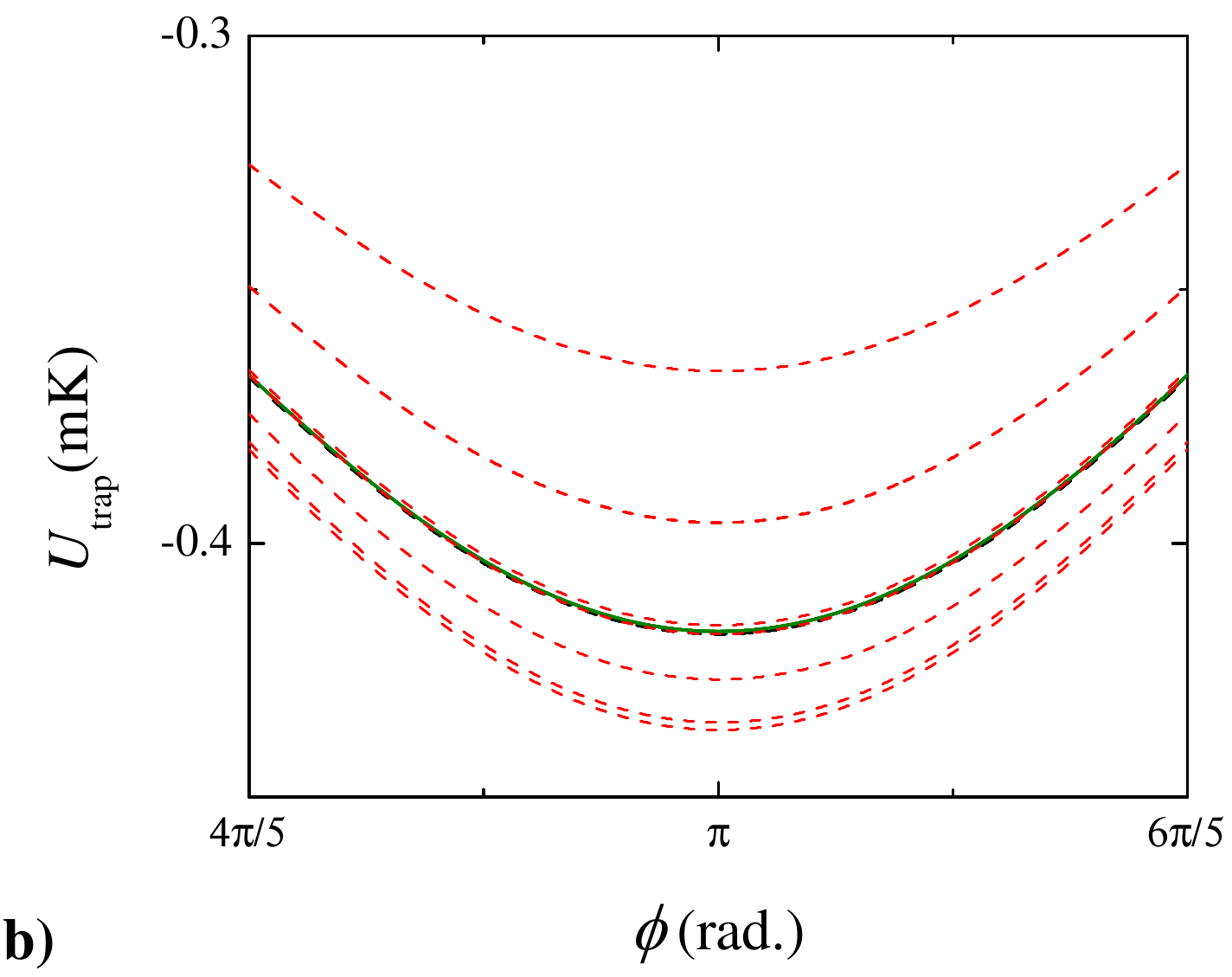}
\caption{\textbf{a)} Azimuthal dependence of the trapping potential of the ground and excited state for the ``magic compensated'' trap (Fig. \ref{fig:configs}d) with the parameters used in Fig. \ref{fig:magic_comp_radial}, for $r-a=210$ nm. \textbf{b)} Expanded view of \textbf{a)} around a trap minimum. The effect of the compensation beam in Fig. \ref{fig:configs}d is to suppress the vector shifts and to reduce the ground-state splittings $\delta\nu_{\phi}$ in the transverse plane for $\phi\neq0,\ \pi$.}%
\label{fig:magic_comp_azimuthal}%
\end{figure}

The azimuthal confinement of the atoms is shown by Fig. \ref{fig:magic_comp_azimuthal}. In contrast to the configurations shown in Figs. \ref{fig:configs}b and \ref{fig:configs}c, the inhomogeneous Zeeman broadening from the ellipticity of $\mathbf{E}_\mathrm{blue}$ is greatly reduced thanks to the compensation scheme of Fig. \ref{fig:configs}d. It is non-zero, however, as expressed by Eq. \eqref{eq:correction}. The remaining splitting of the $F=4$ ground state is $\delta_{\nu}\approx 700\ \mathrm{Hz}$, limiting the coherence time to $\tau \lesssim 1/\delta_{\nu}=1.4 \ \mathrm{ms}$.

In the case of perfect cancellation of the vector shift with $\delta_{fb}=0$, a residual non-zero ground state splitting $\delta\nu_{\phi}$ would still arise from the different scalar dynamic polarizabilities of the $6S_{1/2}$ $F=3$ and $F=4$ ground states \cite{Rosenbusch2009a}. For atoms in their azimuthal motional ground state $|n\rangle_{\phi}=|0\rangle_{\phi}$, the single-atom distribution half-width is $\sigma_{r_{\phi}}  \simeq 30 \ \mathrm{nm}$ (or $\sigma_{\phi}\simeq 4^{\circ}$) with azimuthal trap frequency $\nu_{\mathrm{trap}_{\phi}}\simeq 44\ \mathrm{kHz}$ obtained from a harmonic fit of the potential around $\phi=\pi$. We estimate a spin-wave coherence time $\tau_m=1/\Delta\left(\delta\nu_{\phi}\right)\leq 30$ ms, coming from the spread $\Delta\left(\delta\nu_{\phi}\right)=\delta\nu_{\phi}\left(\phi=\pi\right)-\delta\nu_{\phi}\left(\phi=\pi+\sigma_{\phi}\right)\approx30\ \mathrm{Hz}$ of the atomic ground states for the $F=3\rightarrow F=4$ transition frequency. 

We note that the longest achievable coherence time in the ``magic compensated'' adiabatic potential in the absence of ground-state splitting $\delta \nu_{\phi}$ would be limited by spontaneous Raman scattering driven by the trapping beams \cite{McKeever2003}.

\begin{figure}[h]\centering%
\includegraphics[width=0.45\columnwidth]{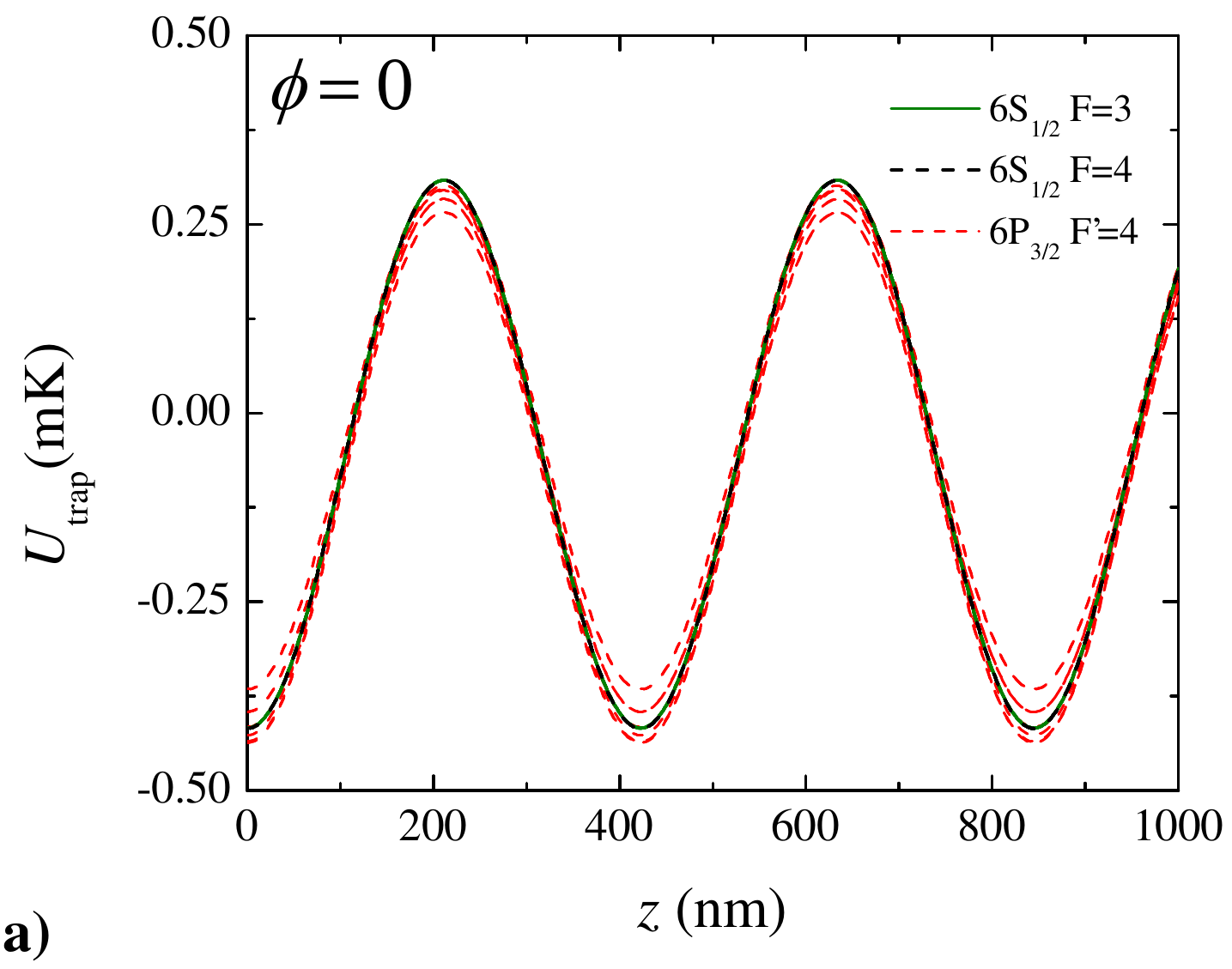}%
\includegraphics[width=0.45\columnwidth]{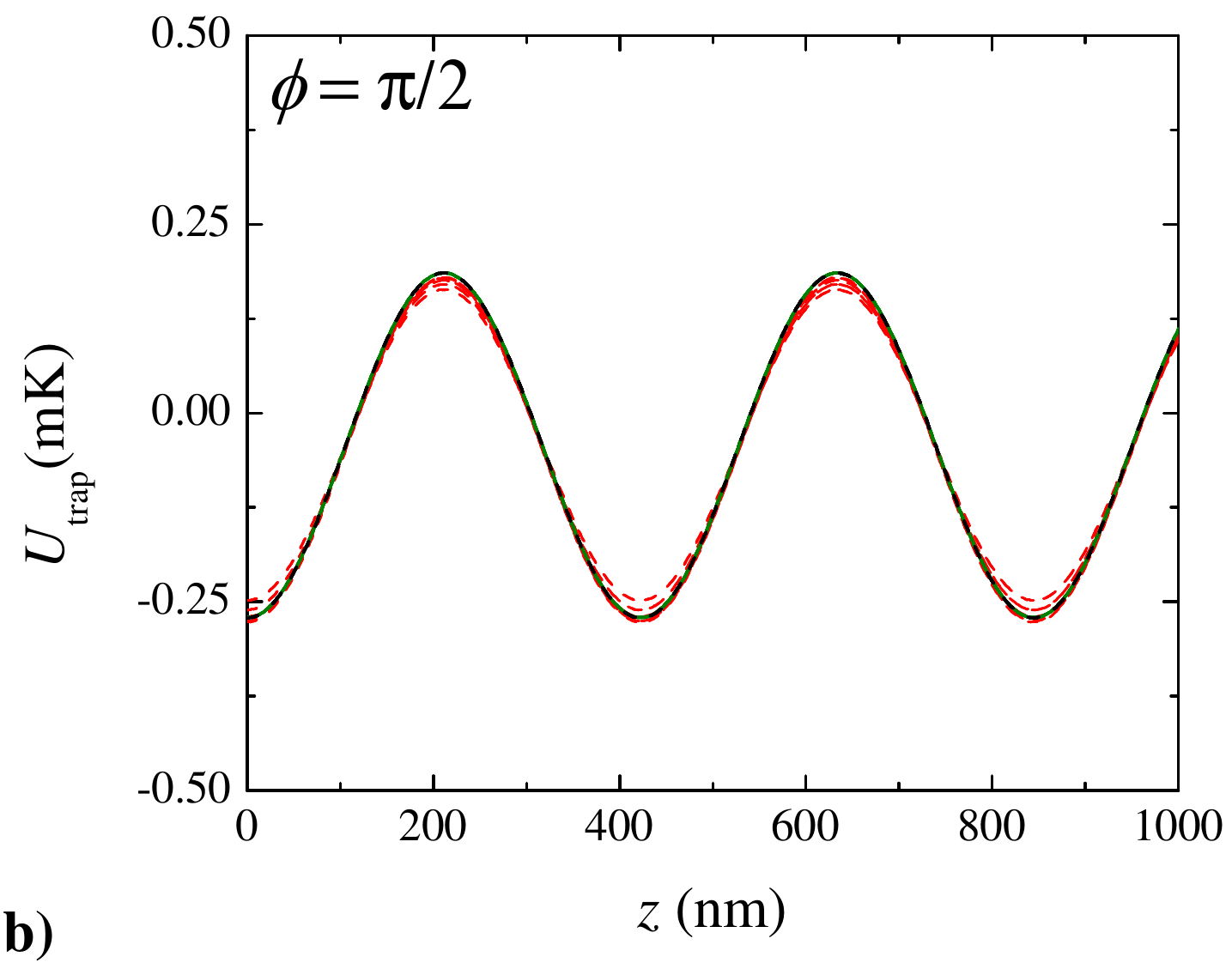}%
\caption{Axial dependence of the trapping potentials of the ground and excited states for the ``magic compensated'' trap (Fig. \ref{fig:configs}d) with the parameters used in Fig. \ref{fig:magic_comp_radial}. \textbf{a)} Longitudinal potential along $\phi=0$ for $r-a=210\ \mathrm{nm}$. \textbf{b)} Longitudinal potential along $\phi=\pi/2$ for $r-a=210\ \mathrm{nm}$.}%
\label{fig:magic_comp_axial}%
\end{figure}

Finally, we also plot the axial potentials in Fig. \ref{fig:magic_comp_axial}, showing the confinement for both the ground and excited states. 

Thanks to the use of the magic wavelengths, the excited states are trapped in all directions. This results in greatly suppressed dipole-force fluctuations, allowing for on-resonance driving of the optical transition.

\section{Conclusion}
We have proposed and analyzed in detail a scheme for a state-insensitive optical nanofiber trap which utilizes realistic experimental parameters. The ``magic'' trapping wavelengths of 937 nm and 687 nm for Cs atoms are readily available using semiconductor laser sources, and require a reasonable power for trapping. Tapered optical fibers can be made with sub-wavelength diameters and high transmission, as has been demonstrated experimentally \cite{Stiebeiner2010, Aoki2010}. In Ref. \cite{exp2011} we describe an experiment for trapping Cs using Fig. \ref{fig:configs}d configuration and explicitly demonstrate features of the ``compensated magic'' trap for a nanofiber.

Furthermore, extension to other nanoscopic dielectric waveguides \cite{Eichenfield2009} would make evanescent optical trapping possible in integrated hybrid quantum devices \cite{Wallquist2009}. It is worthwhile to note that the compensation scheme of the vector shift would work at any wavelength, increasing the ground state coherence time in a straightforward manner.
 
The proposed ``magic compensated'' scheme allows for in-trap resonant processes, and leads to increased ground state coherence time relative to uncompensated schemes. These properties will make quantum-state engineering more feasible in such a trap, allowing for a wide range of experiments including creating quantum memories, coupling of single atoms and ensembles to optical or mechanical resonators, and studying 1-D spin chains.

\section{Acknowledgment}
We gratefully acknowledge interactions with E. S. Polzik and M. Pototschnig. Funding is provided by the Institute for Quantum Information and Matter, an NSF Physics Frontier Center with support of the Gordon and Betty Moore Foundation, by NSF Grant $\#$ PHY0652914, by the DoD NSSEFF program, by the AFOSR MURI for Quantum Memories, and by Northrop Grumman Aerospace Systems. Akihisa Goban is supported by the Nakajima Foundation.

\section*{References}
\bibliographystyle{unsrt}
\bibliography{biblio}

\begin{thebibliography}{10}

\bibitem{Kimble2008}
H.~J. Kimble.
\newblock The quantum internet.
\newblock {\em Nature}, 453(7198):1023--1030, June 2008.

\bibitem{Renn1995}
M.~J. Renn, D.~Montgomery, O.~Vdovin, D.~Z. Anderson, C.~E. Wieman, and E.~A.
  Cornell.
\newblock Laser-guided atoms in hollow-core optical fibers.
\newblock {\em Phys. Rev. Lett.}, 75(18):3253--3256, October 1995.

\bibitem{Ito1996}
H.~Ito, T.~Nakata, K.~Sakaki, M.~Ohtsu, K.~I. Lee, and W.~Jhe.
\newblock Laser spectroscopy of atoms guided by evanescent waves in
  micron-sized hollow optical fibers.
\newblock {\em Phys. Rev. Lett.}, 76(24):4500--4503, June 1996.

\bibitem{Christensen2008}
Caleb~A. Christensen, Sebastian Will, Michele Saba, Gyu-Boong Jo, Yong-Il Shin,
  Wolfgang Ketterle, and David Pritchard.
\newblock Trapping of ultracold atoms in a hollow-core photonic crystal fiber.
\newblock {\em Phys. Rev. A}, 78(3):033429, 2008.

\bibitem{Bajcsy2009}
M.~Bajcsy, S.~Hofferberth, V.~Balic, T.~Peyronel, M.~Hafezi, A.~S. Zibrov,
  V.~Vuletic, and M.~D. Lukin.
\newblock Efficient all-optical switching using slow light within a hollow
  fiber.
\newblock {\em Phys. Rev. Lett.}, 102(20):203902, May 2009.

\bibitem{Vetsch2010}
E.~Vetsch, D.~Reitz, G.~Sague, R.~Schmidt, S.~T. Dawkins, and
  A.~Rauschenbeutel.
\newblock Optical interface created by {Laser-Cooled} atoms trapped in the
  evanescent field surrounding an optical nanofiber.
\newblock {\em Physical Review Letters}, 104(20):203603, May 2010.

\bibitem{Kien2004a}
Fam~Le Kien, J.Q. Liang, K.~Hakuta, and V.I. Balykin.
\newblock Field intensity distributions and polarization orientations in a
  vacuum-clad subwavelength-diameter optical fiber.
\newblock {\em Optics Communications}, 242(4-6):445--455, December 2004.

\bibitem{Nayak2007}
K.~P. Nayak, P.~N. Melentiev, M.~Morinaga, Fam~Le Kien, V.~I. Balykin, and
  K.~Hakuta.
\newblock Optical nanofiber as an efficient tool for manipulating and probing
  atomic fluorescence.
\newblock {\em Opt. Express}, 15(9):5431--5438, April 2007.

\bibitem{Sague2007}
G.~Sague, E.~Vetsch, W.~Alt, D.~Meschede, and A.~Rauschenbeutel.
\newblock {Cold-Atom} physics using ultrathin optical fibers: {Light-Induced}
  dipole forces and surface interactions.
\newblock {\em Physical Review Letters}, 99(16):163602, October 2007.

\bibitem{Balykin1988}
V.~I. Balykin, V.~S. Letokhov, Yu.~B. Ovchinnikov, and A.~I. Sidorov.
\newblock Quantum-state-selective mirror reflection of atoms by laser light.
\newblock {\em Phys. Rev. Lett.}, 60(21):2137--2140, May 1988.

\bibitem{Cronin2009}
Alexander~D. Cronin, J{\"o}rg Schmiedmayer, and David~E. Pritchard.
\newblock Optics and interferometry with atoms and molecules.
\newblock {\em Rev. Mod. Phys.}, 81(3):1051--1129, July 2009.

\bibitem{Ovchinnikov1997}
Yu.~B. Ovchinnikov, I.~Manek, and R.~Grimm.
\newblock Surface trap for {Cs} atoms based on evanescent-wave cooling.
\newblock {\em Phys. Rev. Lett.}, 79(12):2225--2228, 1997.

\bibitem{Rychtarik2004}
D.~Rychtarik, B.~Engeser, H.-C. Nägerl, and R.~Grimm.
\newblock Two-dimensional {Bose-Einstein} condensate in an optical surface
  trap.
\newblock {\em Phys. Rev. Lett.}, 92(17):173003, April 2004.

\bibitem{Bakr2010}
W.~S. Bakr, A.~Peng, M.~E. Tai, R.~Ma, J.~Simon, J.~I. Gillen, S.~Folling,
  L.~Pollet, and M.~Greiner.
\newblock Probing the {Superfluid-to-Mott} insulator transition at the
  {Single-Atom} level.
\newblock {\em Science}, 329(5991):547--550, July 2010.

\bibitem{Boozer2007}
A.~D. Boozer, A.~Boca, R.~Miller, T.~E. Northup, and H.~J. Kimble.
\newblock Reversible state transfer between light and a single trapped atom.
\newblock {\em Phys. Rev. Lett.}, 98(19):193601, May 2007.

\bibitem{Choi2008}
K.~S. Choi, H.~Deng, J.~Laurat, and H.~J. Kimble.
\newblock Mapping photonic entanglement into and out of a quantum memory.
\newblock {\em Nature}, 452(7183):67--71, March 2008.

\bibitem{Hammerer2010}
Klemens Hammerer, Anders~S. Sørensen, and Eugene~S. Polzik.
\newblock Quantum interface between light and atomic ensembles.
\newblock {\em Rev. Mod. Phys.}, 82(2):1041--1092, April 2010.

\bibitem{Sangouard2011}
Nicolas Sangouard, Christoph Simon, Hugues de~Riedmatten, and Nicolas Gisin.
\newblock Quantum repeaters based on atomic ensembles and linear optics.
\newblock {\em Rev. Mod. Phys.}, 83(1):33--80, March 2011.

\bibitem{Aoki2006}
Takao Aoki, Barak Dayan, E.~Wilcut, W.~P. Bowen, A.~S. Parkins, T.~J.
  Kippenberg, K.~J. Vahala, and H.~J. Kimble.
\newblock Observation of strong coupling between one atom and a monolithic
  microresonator.
\newblock {\em Nature}, 443(7112):671--674, October 2006.

\bibitem{Alton2011}
D.~J. Alton, N.~P. Stern, Takao Aoki, H.~Lee, E.~Ostby, K.~J. Vahala, and H.~J.
  Kimble.
\newblock Strong interactions of single atoms and photons near a dielectric
  boundary.
\newblock {\em Nat Phys}, 7(2):159--165, February 2011.

\bibitem{Stern2011}
N.~P. Stern, D.~J. Alton, and H.~J. Kimble.
\newblock Simulations of atomic trajectories near a dielectric surface.
\newblock {\em New J. Phys.}, 13(8):085004, August 2011.

\bibitem{Barclay2006}
Paul~E. Barclay, Kartik Srinivasan, Oskar Painter, Benjamin Lev, and Hideo
  Mabuchi.
\newblock Integration of fiber-coupled {high-Q} {SiN}$_x$ microdisks with atom
  chips.
\newblock {\em Applied Physics Letters}, 89:131108, 2006.

\bibitem{Colombe2007}
Yves Colombe, Tilo Steinmetz, Guilhem Dubois, Felix Linke, David Hunger, and
  Jakob Reichel.
\newblock Strong atom-field coupling for bose-einstein condensates in an
  optical cavity on a chip.
\newblock {\em Nature}, 450(7167):272--276, November 2007.

\bibitem{Trupke2007}
M.~Trupke, J.~Goldwin, B.~Darqui{\'e}, G.~Dutier, S.~Eriksson, J.~Ashmore, and
  E.~A. Hinds.
\newblock Atom detection and photon production in a scalable, open, optical
  microcavity.
\newblock {\em Phys. Rev. Lett.}, 99(6):063601, 2007.

\bibitem{Zoubi2010}
Hashem Zoubi and Helmut Ritsch.
\newblock Hybrid quantum system of a nanofiber mode coupled to two chains of
  optically trapped atoms.
\newblock {\em New J. Phys.}, 12(10):103014, 2010.

\bibitem{Kitagawa2011}
Takuya Kitagawa, Matthew~A Broome, Alessandro Fedrizzi, Mark~S Rudner, Erez
  Berg, Ivan Kassal, Al\'an Aspuru-Guzik, Eugene Demler, and Andrew~G White.
\newblock Observation of topologically protected bound states in a one
  dimensional photonic system.
\newblock arXiv:1105.5334, May 2011.

\bibitem{Shen2005}
J.~T. Shen and Shanhui Fan.
\newblock Coherent photon transport from spontaneous emission in
  one-dimensional waveguides.
\newblock {\em Opt. Lett.}, 30(15):2001--2003, 2005.

\bibitem{Chang2007}
Darrick~E. Chang, Anders~S. Sorensen, Eugene~A. Demler, and Mikhail~D. Lukin.
\newblock A single-photon transistor using nanoscale surface plasmons.
\newblock {\em Nat Phys}, 3(11):807--812, November 2007.

\bibitem{Corwin1999}
K.~L. Corwin, S.~J.~M. Kuppens, D.~Cho, and C.~E. Wieman.
\newblock Spin-polarized atoms in a circularly polarized optical dipole trap.
\newblock {\em Phys. Rev. Lett.}, 83(7):1311--1314, 1999.

\bibitem{Ye2008}
Jun Ye, H.~J. Kimble, and Hidetoshi Katori.
\newblock Quantum state engineering and precision metrology using
  state-insensitive light traps.
\newblock {\em Science}, 320(5884):1734--1738, June 2008.

\bibitem{Kimble1999}
H.~J. Kimble.
\newblock In R.~Blatt, J.~Eschner, D.~Leibfried, and F.~Schmidt-Kaler, editors,
  {\em {Proceedings of the XIV International Conference on Laser
  Spectroscopy}}, volume XIV, pages 80 -- 89. World Scientific, Innsbruck,
  Austria, 1999.

\bibitem{Katori1999}
Hidetoshi Katori, Tetsuya Ido, and Makoto Kuwata-Gonokami.
\newblock Optimal design of dipole potentials for efficient loading of {Sr}
  atoms.
\newblock {\em J. Phys. Soc. Jpn.}, 68(8):2479--2482, 1999.

\bibitem{Ido2000}
Tetsuya Ido, Yoshitomo Isoya, and Hidetoshi Katori.
\newblock Optical-dipole trapping of {Sr} atoms at a high phase-space density.
\newblock {\em Phys. Rev. A}, 61(6):061403, May 2000.

\bibitem{McKeever2003}
J.~McKeever, J.~R. Buck, A.~D. Boozer, A.~Kuzmich, H.-C. Nägerl, D.~M.
  Stamper-Kurn, and H.~J. Kimble.
\newblock State-insensitive cooling and trapping of single atoms in an optical
  cavity.
\newblock {\em Phys. Rev. Lett.}, 90(13):133602, April 2003.

\bibitem{Kien2005a}
Fam~Le Kien, Victor~I. Balykin, and Kohzo Hakuta.
\newblock {State-Insensitive} trapping and guiding of {Cesium} atoms using a
  {Two-Color} evanescent field around a {Subwavelength-Diameter} fiber.
\newblock {\em Journal of the Physics Society Japan}, 74(3):910--917, 2005.

\bibitem{Dupont-Roc1967}
J.~Dupont-Roc, N.~Polonsky, C.~Cohen-Tannoudji, and A.~Kastler.
\newblock Lifting of a {Zeeman} degeneracy by interaction with a light beam.
\newblock {\em Physics Letters A}, 25(2):87--88, July 1967.

\bibitem{Felinto2005}
D.~Felinto, C.~W. Chou, H.~de~Riedmatten, S.~V. Polyakov, and H.~J. Kimble.
\newblock Control of decoherence in the generation of photon pairs from atomic
  ensembles.
\newblock {\em Phys. Rev. A}, 72(5):053809, November 2005.

\bibitem{Balykin2004}
V.~I. Balykin, K.~Hakuta, Fam~Le Kien, J.~Q. Liang, and M.~Morinaga.
\newblock Atom trapping and guiding with a subwavelength-diameter optical
  fiber.
\newblock {\em Physical Review A}, 70(1):011401, July 2004.

\bibitem{Dawkins2011}
S.~T. Dawkins, R.~Mitsch, D.~Reitz, E.~Vetsch, and A.~Rauschenbeutel.
\newblock Dispersive optical interface based on nanofiber-trapped atoms.
\newblock ArXiv 1108.2469, August 2011.

\bibitem{Deutsch2010a}
Ivan~H. Deutsch and Poul~S. Jessen.
\newblock Quantum control and measurement of atomic spins in polarization
  spectroscopy.
\newblock {\em Optics Communications}, 283(5):681--694, March 2010.

\bibitem{Geremia2006}
J.~M. Geremia, John~K. Stockton, and Hideo Mabuchi.
\newblock Tensor polarizability and dispersive quantum measurement of
  multilevel atoms.
\newblock {\em Phys. Rev. A}, 73(4):042112, April 2006.

\bibitem{Deutsch1998}
Ivan~H. Deutsch and Poul~S. Jessen.
\newblock Quantum-state control in optical lattices.
\newblock {\em Phys. Rev. A}, 57(3):1972--1986, March 1998.

\bibitem{Chicireanu2011}
R.~Chicireanu, K.~D. Nelson, S.~Olmschenk, N.~Lundblad, A.~Derevianko, and
  J.~V. Porto.
\newblock Differential light-shift cancellation in a magnetic-field-insensitive
  transition of $^{87}${Rb}.
\newblock {\em Phys. Rev. Lett.}, 106(6):063002, February 2011.

\bibitem{Tong2004}
Limin Tong, Jingyi Lou, and Eric Mazur.
\newblock Single-mode guiding properties of subwavelength-diameter silica and
  silicon wire waveguides.
\newblock {\em Opt. Express}, 12(6):1025--1035, March 2004.

\bibitem{Sague2008}
G~Sague, A~Baade, and A~Rauschenbeutel.
\newblock Blue-detuned evanescent field surface traps for neutral atoms based
  on mode interference in ultrathin optical fibres.
\newblock {\em New Journal of Physics}, 10(11):113008, 2008.

\bibitem{Ashraf2001}
M.~M. Ashraf.
\newblock The effects of phase and {Stark} shift on two-photon process.
\newblock {\em J. Opt. B: Quantum Semiclass. Opt.}, 3(2):39--43, April 2001.

\bibitem{Savard1997}
T.~A. Savard, K.~M. O'Hara, and J.~E. Thomas.
\newblock Laser-noise-induced heating in far-off resonance optical traps.
\newblock {\em Phys. Rev. A}, 56(2):R1095--R1098, 1997.

\bibitem{Katori2003}
Hidetoshi Katori, Masao Takamoto, V.~G. Pal'chikov, and V.~D. Ovsiannikov.
\newblock Ultrastable optical clock with neutral atoms in an engineered light
  shift trap.
\newblock {\em Phys. Rev. Lett.}, 91(17):173005, October 2003.

\bibitem{Arora2007}
Bindiya Arora, M.~S. Safronova, and Charles~W. Clark.
\newblock Magic wavelengths for the np-ns transitions in alkali-metal atoms.
\newblock {\em Phys. Rev. A}, 76(5):052509, November 2007.

\bibitem{DiBerardino1998}
D.~DiBerardino, C.~E. Tanner, and A.~Sieradzan.
\newblock Lifetime measurements of cesium $5d ^{2}d_{5/2,3/2}$ and $11s
  ^{2}s_{1/2}$ states using pulsed-laser excitation.
\newblock {\em Phys. Rev. A}, 57(6):4204--4211, June 1998.

\bibitem{Sukenik1993}
C.~I. Sukenik, M.~G. Boshier, D.~Cho, V.~Sandoghdar, and E.~A. Hinds.
\newblock Measurement of the {Casimir-Polder} force.
\newblock {\em Phys. Rev. Lett.}, 70(5):560--563, February 1993.

\bibitem{Bordag2001}
M.~Bordag, U.~Mohideen, and V.~M. Mostepanenko.
\newblock New developments in the{ Casimir} effect.
\newblock {\em Physics Reports}, 353(1-3):1--205, October 2001.

\bibitem{Sandoghdar1992}
V.~Sandoghdar, C.~I. Sukenik, E.~A. Hinds, and Serge Haroche.
\newblock Direct measurement of the {van der Waals} interaction between an atom
  and its images in a micron-sized cavity.
\newblock {\em Phys. Rev. Lett.}, 68(23):3432--3435, June 1992.

\bibitem{Hinds1997}
E.~A. Hinds, K.~S. Lai, and M.~Schnell.
\newblock Atoms in micron-sized metallic and dielectric waveguides.
\newblock {\em Philosophical Transactions of the Royal Society of London.
  Series A: Mathematical, Physical and Engineering Sciences},
  355(1733):2353--2365, December 1997.

\bibitem{Emig2006}
T.~Emig, R.~L. Jaffe, M.~Kardar, and A.~Scardicchio.
\newblock Casimir interaction between a plate and a cylinder.
\newblock {\em Phys. Rev. Lett.}, 96(8):080403, March 2006.

\bibitem{Laliotis2007}
A.~Laliotis.
\newblock Testing the distance-dependence of the van der waals interaction
  between an atom and a surface through spectroscopy in a vapor nanocell.
\newblock {\em Proc. SPIE}, 6604(1):660406, 2007.

\bibitem{Radnaev2010}
A.~G. Radnaev, Y.~O. Dudin, R.~Zhao, H.~H. Jen, S.~D. Jenkins, A.~Kuzmich, and
  T.~A.~B. Kennedy.
\newblock A quantum memory with telecom-wavelength conversion.
\newblock {\em Nat. Phys.}, 6(11):894--899, November 2010.

\bibitem{Vetsch2010a}
E.~Vetsch.
\newblock {\em Optical Interface Based on a Nanofiber Atom-Trap}.
\newblock PhD thesis, {Johannes Gutenberg-Universit\"at in Mainz}, 2010.

\bibitem{Rosenbusch2009a}
P.~Rosenbusch, S.~Ghezali, V.~A. Dzuba, V.~V. Flambaum, K.~Beloy, and
  A.~Derevianko.
\newblock {ac Stark shift of the Cs microwave atomic clock transitions}.
\newblock {\em Phys. Rev. A}, 79(1):013404, January 2009.

\bibitem{Stiebeiner2010}
Ariane Stiebeiner, Ruth Garcia-Fernandez, and Arno Rauschenbeutel.
\newblock Design and optimization of broadband tapered optical fibers with a
  nanofiber waist.
\newblock {\em Opt. Express}, 18(22):22677--22685, October 2010.

\bibitem{Aoki2010}
Takao Aoki.
\newblock Fabrication of ultralow-loss tapered optical fibers and microtoroidal
  resonators.
\newblock {\em Jpn. J. Appl. Phys.}, 49(11):118001, November 2010.

\bibitem{exp2011}
In Preparation.

\bibitem{Eichenfield2009}
Matt Eichenfield, Jasper Chan, Ryan~M. Camacho, Kerry~J. Vahala, and Oskar
  Painter.
\newblock Optomechanical crystals.
\newblock {\em Nature}, 462(7269):78--82, November 2009.

\bibitem{Wallquist2009}
M.~Wallquist, K.~Hammerer, P.~Rabl, M.~Lukin, and P.~Zoller.
\newblock Hybrid quantum devices and quantum engineering.
\newblock {\em Phys. Scr.}, T137:014001, December 2009.

\end{thebibliography}
\end{document}